\documentclass[journal]{IEEEtran}
\usepackage{graphicx}
\usepackage{float}
\usepackage{amsmath}
\usepackage{amssymb}
\usepackage{bm}
\usepackage{cuted}
\usepackage{subfiles}
\usepackage{color}
\graphicspath{{../images/}{./images/}}

\usepackage{subcaption}

\usepackage{notoccite}
\usepackage{cite}

\usepackage{algorithm}
\usepackage{algorithmic}
\usepackage[utf8]{inputenc}

\usepackage{amsthm}
\allowdisplaybreaks

\newcommand{\Y}{\bm{Y}}
\newcommand{\X}{\bm{X}}
\newcommand{\W}{\bm{W}}
\newcommand{\Hmat}{\bm{H}}
\newcommand{\F}{\bm{F}}

\newcommand{\Xb}{\bm{X}}
\newcommand{\n}{\bm{n}}
\newcommand{\N}{\bm{N}}
\newcommand{\To}{T_\text{OFDM}}
\newcommand{\ar}{\bm{a_\text{rx}}}
\newcommand{\at}{\bm{a_\text{tx}}}
\newcommand{\thetarn}{\theta_{\text{rx},n}}
\newcommand{\thetatn}{\theta_{\text{tx},n}}
\newcommand{\e}{\text{e}}
\newcommand{\q}{\bm{q}}
\newcommand{\p}{\bm{p}}
\newcommand{\s}{\bm{r}}
\newcommand{\Wa}{\bm{W_a}}
\newcommand{\WaD}{\bm{W}_a^{(D)}}
\newcommand{\WaDTilde}{\Tilde{\bm{W}}_a^{(D)}}
\newcommand{\wan}{\bm{w}_{a,n}}
\newcommand{\wanhat}{\hat{\bm{w}}_{a,n}}
\newcommand{\Fa}{\bm{F}_a}

\newcommand{\FaDTilde}{\Tilde{\bm{F}}_a^{(D)}}
\newcommand{\fan}{\bm{f}_{a,n}}
\newcommand{\phib}{\bm{\phi}}
\newcommand{\Phib}{\bm{\Phi}}

\newcommand{\PhibDTilde}{\Tilde{\bm{\Phi}}^{(D)}}
\newcommand{\SNR}{\text{SNR}}
\newcommand{\G}{\bm{G}}
\newcommand{\V}{\bm{V}}
\newcommand{\U}{\bm{U}}
\newcommand{\Q}{\bm{Q}}
\newcommand{\uvec}[2]{\bm{u}_{#1}^{#2}}
\newcommand{\Sigmab}{\bm{\Sigma}}
\newcommand{\Psib}{\bm{\Psi}}
\newcommand{\Sb}{\bm{S}}
\newcommand{\Db}{\bm{D}}

\newcommand{\vect}[1]{\text{vec}({#1})}
\newcommand{\thetab}{\bm{\theta}}
\newcommand{\T}{N_{\rm T}}

\DeclareMathOperator*{\argmin}{\arg\min}

\title{Multilinear SVD for Millimeter Wave Channel Parameter Estimation}
\author{\IEEEauthorblockN{Macey Ruble and \.{I}smail~G\"uven\c{c}} \\
\IEEEauthorblockA{Dept. Electrical and Computer Engineering, North Carolina State University, Raleigh, NC 27695}\\
Email: \{mcruble,iguvenc\}@ncsu.edu
\vspace{-0.7cm}
}

\begin{document}
\onecolumn 
\pagestyle{empty} 
\begin{center}
  \large\bfseries  
  This work has been submitted to the IEEE for possible publication. Copyright may be transferred without notice, after which this version may no longer be accessible
\end{center}
\twocolumn 
\setcounter{page}{1} 

\maketitle

\begin{abstract}
    Fifth generation (5G) cellular standards are set to utilize millimeter wave (mmWave)
    frequencies, which enable data speeds greater than 10 Gbps and sub-centimeter localization accuracy. These capabilities rely on accurate estimates of the channel parameters, which we define as the angle of arrival, angle of departure, and path distance for each path between the transmitter and receiver. Estimating the channel parameters in a computationally efficient manner poses a challenge because it requires estimation of parameters from a high-dimensional measurement -- particularly for multi-carrier systems since each subcarrier must be estimated separately.  Additionally, channel parameter estimation must be able to handle hybrid beamforming, which uses a combination of digital and analog beamforming to reduce the number of required analog to digital converters.
    This paper introduces a channel parameter estimation technique based on the multilinear singular value decomposition (MSVD), a tensor analog of the singular value decomposition, for massive multiple input multiple output (MIMO) multi-carrier systems with hybrid beamforming. The MSVD tensor estimation approach provides a computationally efficient method and is shown to closely match the Cramer-Rao bound (CRB) of parameter estimates through simulations.  Limitations of channel parameter estimation and communication waveform effects are also studied.    
\end{abstract}


\section{Introduction}
Fifth generation (5G) cellular networks are set to deploy millimeter wave (mmWave) technology with carrier frequencies ranging from $30$~GHz to $300$ GHz \cite{rappaport2013millimeter}.  The small wavelengths associated with mmWave frequencies allow massive multiple-input-multiple-output (MIMO) arrays to be deployed in small spaces \cite{zhu2014demystifying}.  This along with the large available bandwidth enables mmWave technology to reach data rates over $10$~Gbps~\cite{rappaport2013millimeter} while also offering sub-centimeter receiver positioning capability~\cite{shahmansoori2018position}.  

High performance communication and localization for mmWave technology are both dependent on accurate estimates of the channel parameters, which we define as the angle of arrival (AOA), angle of departure (AOD), and total transmitter to receiver path distance for each  significant path.  This dependence is a result of highly directional beams and large path reflection attenuation, which leads to few significant received paths and a sparse channel that is completely characterized by the channel parameters~\cite{deng2014mm}.  Channel estimation for mmWave
is accomplished by transmitting a known training sequence so that the environmental impacts on the channel can be determined and accounted for when unknown data sequences are transmitted \cite{shahmansoori2018position,sohrabi2017hybrid,alkhateeb2014channel}.  The channel parameters from  multiple paths can also be simultaneously utilized to estimate a receiver's position and orientation while also mapping the scatterers in the environment~\cite{shahmansoori2018position,ruble2018}.

5G networks will use orthogonal frequency division multiplexing (OFDM) or OFDM-like waveforms, which enable broadband communication by distributing the transmitted data over many subcarrier frequencies \cite{shahmansoori2018position,guan20175g,sohrabi2017hybrid,hampton2013introduction}.  Channel parameter estimation is particularly challenging for MIMO OFDM systems since the channel is different at each subcarrier.  An antenna with an analog to digital converter (ADC) at every single antenna element becomes too costly for large arrays.  Thus, a channel parameter estimation method must function with hybrid digital/analog beamforming, which uses digital beamforming as well as phased array analog beamforming to reduce the overall number of required ADCs \cite{sohrabi2017hybrid,alkhateeb2014channel}.   Adding to the challenge is that the number of significant received paths is unknown and needs to first be estimated.  Furthermore, a collection of AOA, AOD, and path distance estimates is not a complete solution: a complete solution must also link each estimated channel parameter to particular paths.  

\subsection{Literature Review}\label{sec:litrev}

Massive MIMO channel parameter estimation has received much recent attention in the literature  \cite{deng2014mm,shahmansoori2018position,sohrabi2017hybrid,alkhateeb2014channel,rappaport2015wideband,andrews2016modeling,zhang2018training,suryaprakash2016millimeter,brady2015wideband}, but it still remains a challenge to utilize all of the subcarriers simultaneously for OFDM systems and process the high-dimensional data that streams from the massive MIMO arrays.  These works focus on channel parameter estimation where the transmitter and receiver are all on the same 
plane.  However, these methods may not scale to higher dimensional scenarios when both the azimuthal and elevation angles for AOD and AOA must be considered. 

High accuracy mmWave localization enables a new era of user tracking and augmented reality \cite{deng2014mm,shahmansoori20155g,garcia2017optimal,garcia2017direct,witrisal2016high}. The work in \cite{shahmansoori2018position} estimates channel parameters and shows that the channel parameters for a few paths are sufficient to estimate a receiver's position and orientation. 
Additionally, mmWave non-line of sight (NLOS) paths are not treated as interference, but rather as additional paths that carry useful information \cite{mendrzik2017harnessing,deng2014mm}.  This enables the reflection locations to be estimated from the channel parameters; making simultaneous localization and mapping (SLAM) possible, where the receiver is localized while the environment is mapped in parallel \cite{witrisal2016high,ruble2018}. 


Many other applications are arising that require high dimensional parameter estimation. It has been shown that these problems can be solved efficiently in tensor form \cite{hayes2017low,nion2010tensor,sidiropoulos2017tensor,tensorlab}.  Of particular importance is the multilinear singular value decomposition (MSVD), which is a tensor analogue of the singular value decomposition (SVD) commonly seen in linear algebra.  The MSVD allows computationally efficient parameter estimation if the tensor is low rank and is often used in machine learning~\cite{sidiropoulos2017tensor,tensorlab}.  

MSVD tensor estimation techniques are ideal for channel parameter estimation since an OFDM MIMO receiver measurement can be represented in a low rank tensor form as shown in \cite{zhou2017low}. This is achieved by grouping the receiver measurements into a low rank canonical polyadic decomposition (CPD) tensor form prior to channel parameter estimation.  However, in \cite{zhou2017low} the CPD tensor form has restrictions and requires that no two paths have any channel parameters in common. 

\begin{figure*}[t]
	\includegraphics[width=0.80\textwidth]{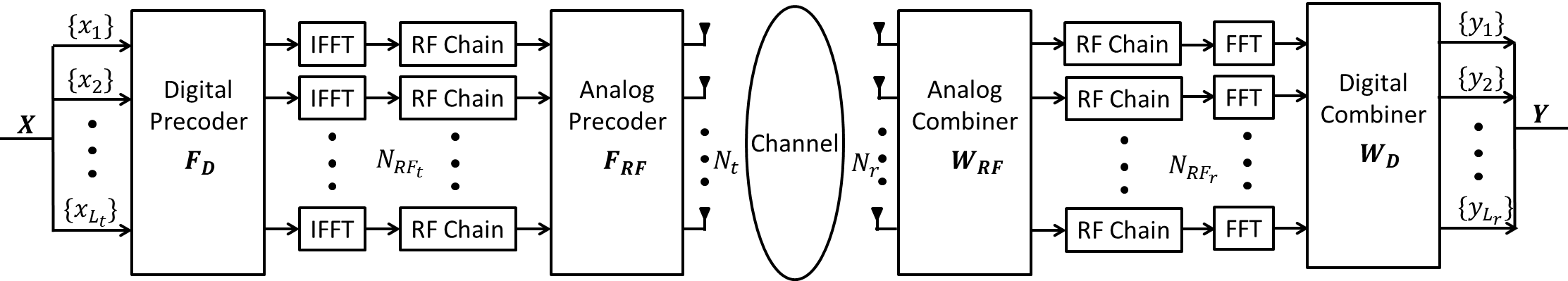}
	\centering
	\caption{MIMO OFDM channel model with hybrid beamforming.}
	\label{fig:OFDMsystem}
\end{figure*}

\subsection{Contributions}
In this work, we propose an alternative method to \cite{zhou2017low} for MIMO OFDM channel parameter estimation that utilizes all of the subcarriers simultaneously by arranging receiver measurements into a low rank Tucker tensor form and employs a MSVD to estimate the channel parameters.  The Tucker tensor form offers multiple advantages over the CPD tensor form.  One reason for this is that the Tucker tensor form is a more natural tensor decomposition for receiver measurements since the path gains and channel parameters are each separated into independent tensor components.  A key advantage to the Tucker form and MSVD estimation is that different paths can share the same channel parameters, which is a restriction in \cite{zhou2017low}.  Furthermore, the Tucker form is not unique nor is it required to be in order to obtain unique estimates of the channel parameters that are correctly linked to path parameters.  

The proposed method first applies the MSVD to the measurement tensor and the number of significant paths are estimated using the multilinear singular values.  Then, the channel parameters are estimated by separating each of the channel parameters into an independent low dimensional sub-problem, making the method computationally efficient. Following this, the low rank structure of the measurement tensor is exploited to link channel parameters to particular paths. 

The Cramer-Rao bound (CRB) performance bound is derived for each of the path parameters. 
Then, the proposed channel parameter estimation method is simulated and shown to closely match the CRB bound.  Simulations of channel parameter estimation are conducted to study a variety of waveform specifications consistent with 5G specifications.  
Our results show that estimation performance for all of the channel parameters is improved by increasing the number of subcarriers, even if the bandwidth is held fixed.

\subsection{Paper Organization}

This paper is organized as follows.  Section~\ref{sec:model} lays out the MIMO OFDM channel model in Tucker tensor form.  Section~\ref{sec:ProbForm} formulates the channel parameter estimation problem from Tucker tensor form receiver measurements.  Then, Section~\ref{sec:MSVD} introduces the MSVD channel parameter estimation technique.  Following this, Section~\ref{sec:wavePar} discusses how the channel parameters can be used for channel estimation and localization followed by how waveform parameters affect channel parameter estimation accuracy.  Subsequently, Section~\ref{sec:SimsAndResults} provides simulation results of channel parameter estimation and compares the results to the CRB bound.  Finally, Section~\ref{sec:Conclusion} provides concluding remarks.

\section{Broadband MIMO OFDM Model}\label{sec:model}
This section covers the model of an OFDM MIMO system.  First, the model is given in a general format representative of hybrid digital/analog beamforming architectures.  Then, it is shown that the MIMO OFDM channel is naturally represented by a low rank Tucker tensor model.  

\subsection{MIMO OFDM Channel Model}
A MIMO OFDM system is considered with $N_{\rm tx}$ transmit antennas and $N_{\rm rx}$ receive antennas (both assumed odd). The data is distributed over a bandwidth $B_\text{OFDM}$ between $N_{\rm s}$ subcarriers at frequencies $f_k = f_c+k/\To$, for $k=0,\dots,N_{\rm s}-1$, where $f_c$ is the carrier frequency and $\To=N_{\rm s} T_s$ is the OFDM symbol duration, where $T_s=1/B_\text{OFDM}$ is the sampling interval.  


Fig.~\ref{fig:OFDMsystem} shows the transmitter, channel, and receiver model, where the data $\X[k] \in \mathbb{C}^{L_{\rm tx} \times \T}$ is divided into $L_{\rm tx} \leq N_{\rm tx}$ data streams of length $\T$ and precoded with a digital precoder $\bm{F_D} \in \mathbb{C}^{L_{\rm tx} \times N_{\text{RF}_{\rm tx}}}$.  A $N_{\rm s}$-point inverse fast Fourier transform (IFFT) is applied to convert the data to the time domain where the output of an IFFT block is
\begin{eqnarray}\label{eq:IFFT}
    s(t) = \sum\limits_{k=0}^{N_{\rm s}-1} x(k) \e^{j2\pi (f_c+m/\To)t}, & \text{for} ~ 0 \leq t \leq T_\text{OFDM}.
\end{eqnarray}
Following this, a cyclic prefix is added to suppress intersymbol interference (ISI), but is not shown in Fig.~\ref{fig:OFDMsystem}.  Then, the transmitter employs an analog precoder $\F_{\bm{\text{RF}}} \in \mathcal{C}^{N_{\text{RF}_{\rm tx}} \times N_{\rm tx}}$ and the signal is transmitted over $N_{\rm tx}$ antennas.  The signal is received by $N_{\rm rx}$ receiving antennas followed by an analog combiner $\W_{\bm{\text{RF}}} \in \mathcal{C}^{N_{\rm rx} \times N_{\text{RF}_{\rm rx}}}$.  A fast Fourier transform (FFT) is then employed as the inverse of the IFFT block before a digital combiner $\W_{\bm{D}} \in \mathcal{C}^{N_{\text{RF}_{\rm rx}} \times L_{\rm rx}}$ converts the signal to $L_{\rm rx} \leq N_{\rm rx}$ data streams to obtain the 
received signal $\Y$.

The model in Fig.~\ref{fig:OFDMsystem} can be represented at each subcarrier by \cite{shahmansoori2018position,sohrabi2017hybrid}:
\begin{eqnarray}\label{eq:meas1}
    \Y[k] = \W[k]^H \Hmat[k]\F[k]\X[k] + \n[k], & \text{for $k=1,\dots,N_{\rm s}$},
\end{eqnarray}
where $k$ is the subcarrier and $\Y[k] \in \mathbb{C}^{L_{\rm rx} \times \T}$ is the received signal at each subcarrier. The matrix
$\F[k]=\F_{\bm{\text{RF}}} \F_{\bm{D}} [k] \in \mathbb{C}^{N_{\rm tx} \times L_{\rm tx}}$ is the precoding matrix, $\Hmat[k] \in \mathbb{C}^{N_{\rm rx} \times N_{\rm tx}}$ is the channel matrix, $\W[k] = \W_{\bm{\text{RF}}} \W_{\bm{D}} [k] \in \mathbb{C}^{L_{\rm rx} \times N_{\rm rx}}$ is the combiner matrix, and $\n[k] \in \mathbb{C}^{N_{\rm rx} \times \T}$ is noise. The precoder and combiner matrices $\F[k]$ and $\W[k]$ can be designed to improve channel estimation \cite{alkhateeb2014channel}.  The signal to noise ratio (SNR) is defined as:
\begin{equation}
    \SNR = \frac{\Big|\Big|\W[k]^H \Hmat[k] \F[k]\X[k]\Big|\Big|^2_F}{\Big|\Big|\n[k]\Big|\Big|^2_F},
\end{equation}
where $||\cdot||_F$ is the Frobenius norm \cite{hampton2013introduction}.  

\begin{figure}[t]
	\includegraphics[width=0.40\textwidth]{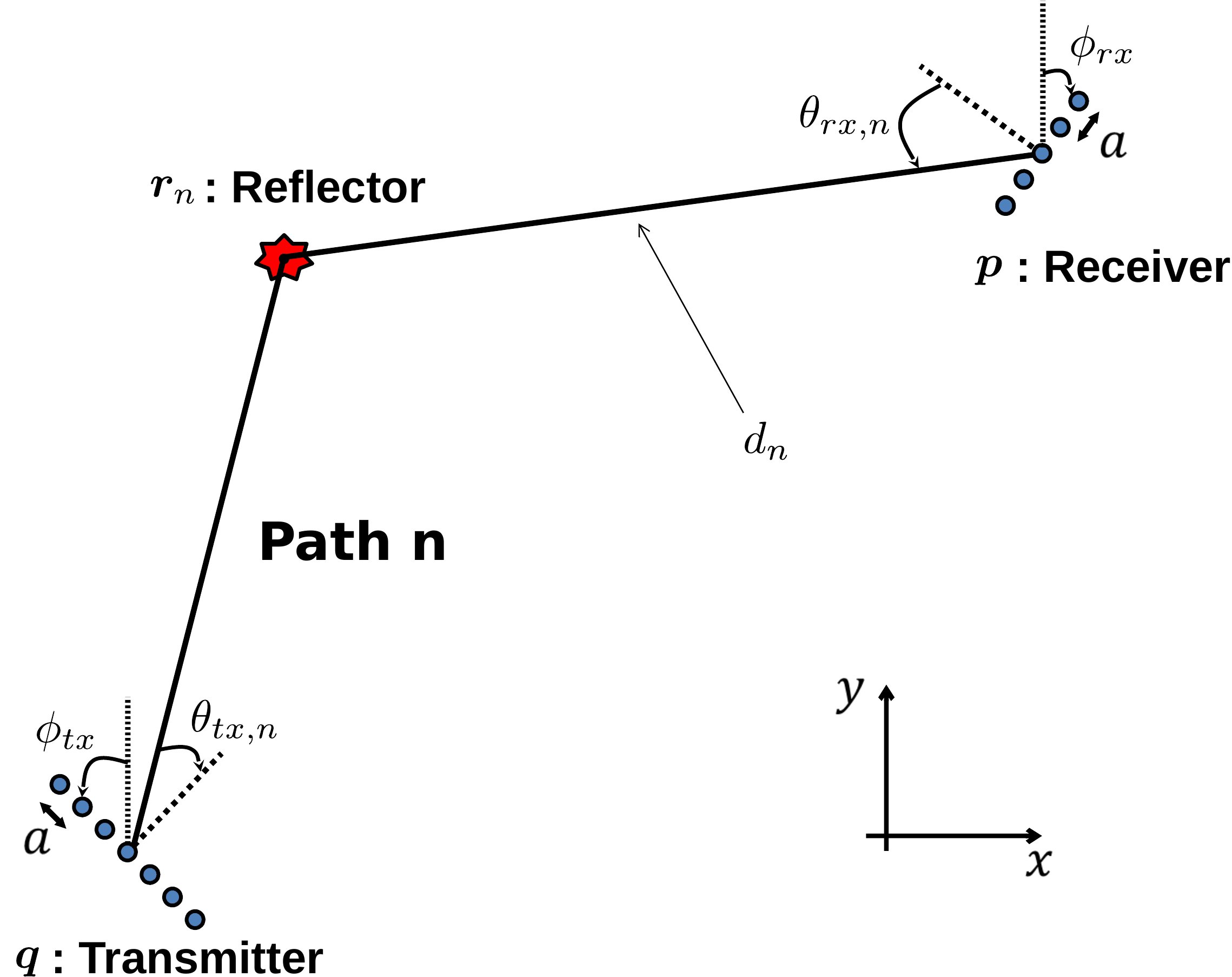}
	\centering
	\caption{Channel parameters for a path between the transmitter array and receiver array.}
	\label{fig:arrayModel}
\end{figure}

The highly directional beams in large scale MIMO systems result in few received significant paths \cite{rappaport2012cellular}. Furthermore, NLOS paths can be assumed single bounce because multiple bounce NLOS paths will have large attenuation and much lower signal strength \cite{andrews2016modeling}.  This results in a sparse channel, which can be expanded in terms of the individual received paths \cite{deng2014mm}.  
We assume $N_{\rm p}$ significant paths are received, where the $n^\text{th}$ path geometry is seen in Fig.~\ref{fig:arrayModel}.  The path begins at the transmitter array located at $\q$ with array orientation angle $\phi_{\rm tx}$, reflects at location $\s$, and ends at a receiver array with location $\p$ and array orientation angle $\phi_{\rm rx}$.  
The path geometry created by the transmitter, reflector, and receiver locations dictate the parameters $\thetatn$: the angle of departure at the transmitter, $\thetarn$: the angle of arrival at the receiver, and $d_n$: the total distance traveled by the path from transmitter to receiver.  It is noted that the path in Fig.~\ref{fig:arrayModel} characterizes both LOS and NLOS paths since a LOS path can be characterized by a reflector anywhere on the line segment between $\q$ and $\p$.
The channel from \eqref{eq:meas1} is expressed in terms of the $N_{\rm p}$ significant paths\cite{shahmansoori2018position,brady2015wideband}:
\begin{equation}\label{eq:channel1}
    \Hmat[k] = \sum\limits_{n=0}^{N_{\rm p}-1} h_n(k) \exp\Bigg\{ \frac{-j 2\pi k \tau_n}{\To} \Bigg\} \ar(\thetarn) \at(\thetatn)^H,
\end{equation}
where $h_n(k)$ is the path loss and $\tau_n=d_n/c$ is the time to travel from the center of the transmitter array to the center of the receiver array, with $c$ as the speed of light. It is assumed that both transmitter and receiver have antenna spacing $a=\lambda_c/2$, where $\lambda_c$ is the wavelength associated with the center frequency.  Then, the beamforming vectors $\ar$ and $\at$ are: 
\begin{align*}
    \ar(\thetarn) &= \big[ \e^{ -j\frac{2\pi a}{\lambda}\big( \frac{N_{\rm rx}-1}{2} \big)\sin(\thetarn)}  \cdots  \e^{ j\frac{2\pi a}{\lambda}\big( \frac{N_{\rm rx}-1}{2} \big)\sin(\thetarn)} \big]^T, \\
    \at(\thetatn) &= \big[ \e^{ -j\frac{2\pi a}{\lambda}\big( \frac{N_{\rm tx}-1}{2} \big)\sin(\thetatn)}  \cdots  \e^{ j\frac{2\pi a}{\lambda}\big( \frac{N_{\rm rx}-1}{2} \big)\sin(\thetatn)} \big]^T.
\end{align*}
For sufficiently long symbols ($\To$) the path loss coefficient is equivalent at each subcarrier \cite{shahmansoori2018position,brady2015wideband}.  Thus, it is assumed that $h_n(k) \approx h_n$ for all $k$. This assumption is discussed in further detail in Section \ref{sec:FreqSel}.

By substituting $t_n=d_n/c$ in \eqref{eq:channel1}, it is seen that besides the path loss $h_n$, the channel is completely characterized by the channel parameters $\thetatn, \thetarn$, and $d_n$:
\begin{equation}\label{eq:channel2}
    \Hmat[k] = \sum\limits_{n=0}^{N_{\rm p}-1} h_n \ar(\thetarn) \at(\thetatn)^H \phi(d_n)[k],
\end{equation}
where,
\begin{equation}
    \phi(d_n)[k] = \exp\Bigg\{ \frac{-j 2\pi k d_n}{c ~ \To} \Bigg\}.
\end{equation}

\subsection{Tucker Tensor Form}
To be self-contained, we cover prerequisite tensor knowledge prior to showing that the received signal in a MIMO OFDM system naturally groups into a Tucker tensor form. Our notation is consistent with \cite{sidiropoulos2017tensor}, which can also be referred to for further information.  We first define the tensor product or outer product. The third order tensor product is defined such that tensor elements resulting from the product between the vectors $\bm{a}$, $\bm{b}$, and $\bm{c}$ are $(\bm{a} \circledcirc \bm{b} \circledcirc \bm{c})(i_1,i_2,i_3) = \bm{a}(i_1) \bm{b}(i_2) \bm{c}(i_3)$.  Tensor products of higher dimension follow similarly.

\begin{figure}[t]
	\includegraphics[width=0.45\textwidth]{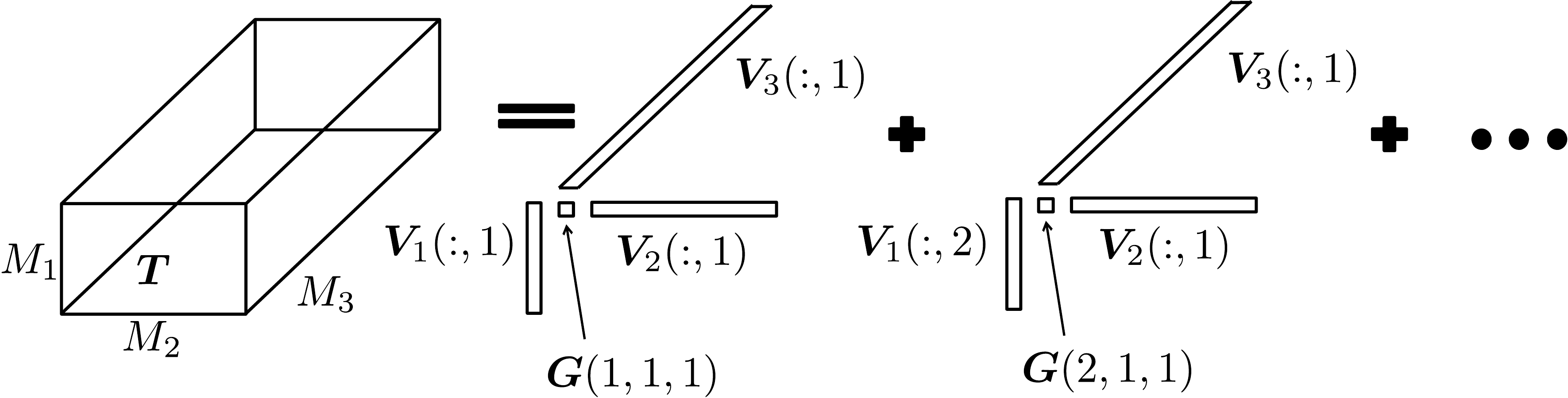}
	\centering
	\caption{Vector-wise outer product view of Tucker model.}
	\label{fig:Tucker1}
\end{figure}

\begin{figure}[t]
	\includegraphics[width=0.45\textwidth]{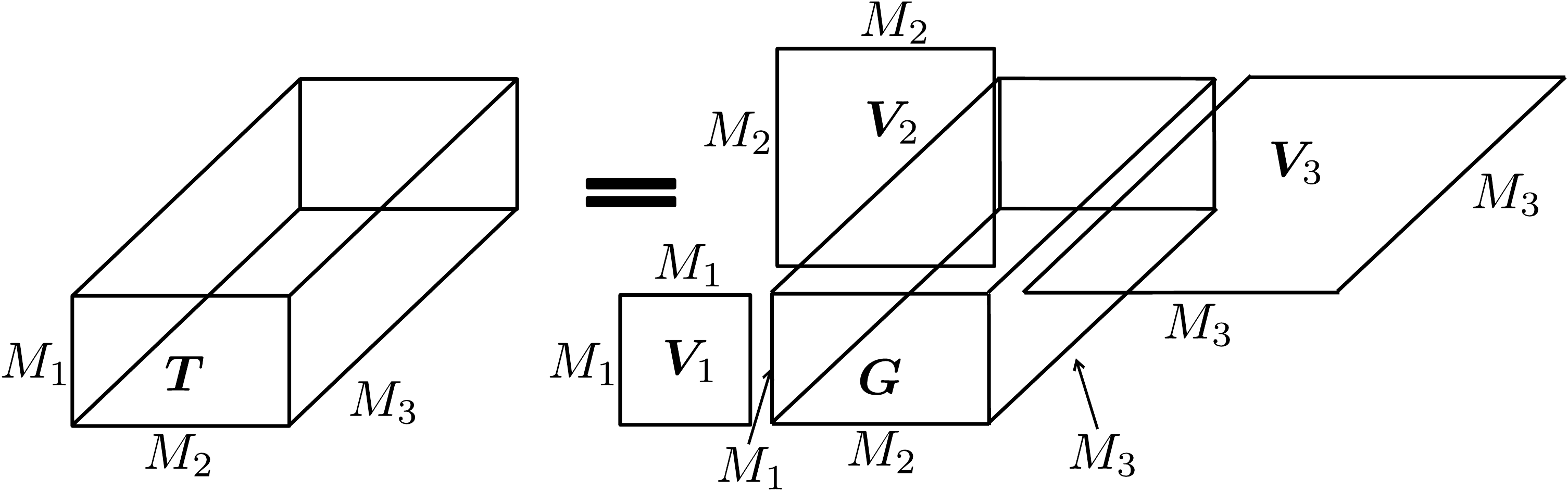}
	\centering
	\caption{Full Tucker tensor form.}
	\label{fig:Tucker2}
\end{figure}

Any tensor can be represented in Tucker form.  For a third order $M_1 \times M_2 \times M_3$ tensor $\bm{T}$, the Tucker form is \cite{sidiropoulos2017tensor}:
\begin{equation}\label{eq:Tucker1}
    \bm{T} = \sum\limits_{i_1=1}^{M_1} \sum\limits_{i_2=1}^{M_2} \sum\limits_{i_3=1}^{M_3} \G(i_1,i_2,i_3) ~ \V_1(:,i_1) \circledcirc \V_2(:,i_2) \circledcirc \V_3(:,i_3),
\end{equation}
where $\G$ is a core tensor, and $\V_1,\V_2,\V_3$ are matrices composed of column vectors such that $\V(:,i)$ is used to represent the $i^\text{th}$ column of matrix $\V$. The core element $\G(i_1,i_2,i_3)$ corresponds with the strength of the interaction between column $i_1$ in $\V_1$, column $i_2$ in $\V_2$, and column $i_3$ in $\V_3$.  Fig.~\ref{fig:Tucker1} visualizes $\bm{T}$ as a series of vector tensor products.  The Tucker tensor form in \eqref{eq:Tucker1} can also be written in a shorthand notation as follows:
\begin{equation}\label{eq:Tucker2}
    \bm{T} =  \G ~  \circledcirc_1 \V_1 \circledcirc_2 \V_2 \circledcirc_3 \V_3,
\end{equation}
where $\circledcirc_i$ represents multiplication along the $i^\text{th}$ dimension with the core tensor $\G$. Fig.~\ref{fig:Tucker2} visualizes this form.
Now that the Tucker form has been introduced, the following subsection shows that the MIMO OFDM received signal naturally groups into a Tucker tensor form.

\subsection{The Channel in Tucker Tensor Form}
Channel estimation and localization  are typically performed during a training sequence interval, where the data $\X$ is known \cite{shahmansoori2018position,choi2015compressed}.  With knowledge of the data, the only unknowns are the channel parameters.  Without loss of generality, we let $\X$ be all zero, besides ones on the diagonal. Substituting $\X$ and the channel representation from \eqref{eq:channel2} into \eqref{eq:meas1}, the received signal for each subcarrier is
\begin{eqnarray}\label{eq:meas2}
    \Y[k] = \sum\limits_{n=0}^{N_{\rm p}-1} h_n \wan(\thetarn) \fan(\thetatn)  \phi(d_n)[k] + \n[k],
\end{eqnarray}
for $k=1,\dots,N_{\rm s}$ where
\begin{equation}\label{eq:wan}
    \wan(\thetarn) = \W^H \ar(\thetarn),
\end{equation}
and
\begin{equation}
    \fan(\thetatn) = \at(\thetarn)^H \F.
\end{equation}

A vector is created from the terms $\phi(d_n)[k]$ that contains $\{\phi(d_n)[k]\}_{k=1}^{N_{\rm s}}$ across all subcarrier frequencies: 
\begin{equation}\label{eq:phin}
    \phib_n(d_n) = \begin{bmatrix}
    1 & \e^{ \frac{-j 2\pi d_n}{c ~ \To} } & \cdots & \e^{ \frac{-j 2\pi (N_{\rm s}-1) d_n}{c ~ \To} }
    \end{bmatrix}^T.
\end{equation}
Then, the measurement across all subcarrier frequencies can be constructed as a third-order tensor with dimensions $L_{\rm rx} \times \T \times N_{\rm s}$ as follows:
\begin{equation}\label{eq:Tuckermeas}
    \Y = \sum\limits_{n=0}^{N_{\rm p}-1} h_n \wan \circledcirc \fan^T \circledcirc \phib_n + \n,
\end{equation}
which is now in Tucker form with rank $N_{\rm p}$ as seen in \eqref{eq:Tucker1}, where $\n \in \mathbb{C}^{L_{\rm rx} \times \T \times N_{\rm s}}$ is the noise over all subcarriers.  It is noted that the dependencies of the vectors on the parameters $\thetarn, \thetatn$, and $d_n$ have been dropped for simplicity in notation.  
The simplified Tucker form from \eqref{eq:Tucker2} is obtained by grouping the vectors $\wan$ into the matrix $\Wa=[\bm{w}_{a,1}~\bm{w}_{a,2}~\dots]$, the vectors $\fan$ into the matrix $\Fa=[\bm{f}_{a,1}^T~\bm{f}_{a,2}^T~\dots]$, and the vectors $\phib_n$ into the matrix $\Phib=[\phib_{1}~\phib_{2}~\dots]$. Then, the measurement tensor is:
\begin{equation}\label{eq:measTuckerFinal}
    \Y = \Psib \circledcirc_1 \Wa \circledcirc_2 \Fa \circledcirc_3 \Phib + \n,
\end{equation}
where $\Psib$ is a core tensor with only $N_{\rm p}$ nonzero elements, corresponding with $h_n$ for each path.

This work focuses on third order tensors because we restrict paths to a plane and do not consider elevation angles for simplicity. The result is a third order measurement tensor where the column space corresponds with path AOA, the row space corresponds with path AOD, and the fiber space corresponds with path distance. A model that allows a three-dimensional path and considers both elevation angles and azimuthal path angles will lead to a five-dimensional measurement tensor. A significant advantage of the Tucker form is that it extends to higher dimensions.  All of the derivations in this work are for three-dimensional tensors, but each step can easily be extended and applied to higher order tensors for path models that also consider elevation angle.

Accurate estimates of the channel parameters $\thetarn$, $\thetatn$, and $d_n$ for $n=0,\dots,N_{\rm p}-1$ enable a number of capabilities for 5G networks.  The most obvious capability is channel estimation, since $\Hmat[k]$ can be reconstructed from the channel parameters for each subcarrier \cite{alkhateeb2014channel}. Additionally, the channel parameters can be exploited to estimate the receiver position and NLOS path reflection locations \cite{ruble2018}.  Thus, receiver localization and environmental mapping can be  achieved with the knowledge of these parameters, where environmental mapping is obtained when the NLOS path reflection locations are estimated over several measurements.  The estimate of $d_n$ from $\phib_n$ can also be used to estimate the time delay $\tau_n$ between the transmitter and receiver, which can be used to assist synchronization between the transmitter and receiver \cite{koivisto2016joint}.


\section{Problem Formulation for mmWave Channel Parameter Estimation}\label{sec:ProbForm}
We are interested in estimating the channel parameters $\thetarn$, $\thetatn$, and $d_n$ for $n=0,\dots,N_{\rm p}-1$ as well as the path gain $h_n$ from the tensor form of the receiver measurement $\Y$ of a training sequence as in \eqref{eq:measTuckerFinal}.  This is accomplished by estimating $\Psib,\Wa,\Fa$, and $\Phib$, which is the main goal of the rest of this paper.  A number of challenges make this a difficult task.  For example, the number of paths $N_{\rm p}$ is unknown and must also be estimated.  Additionally, this is a high dimensional problem and the channel parameters must be estimated jointly.

Mathematically, since the measurement tensor is low rank, the channel parameter estimation problem can be posed in the following form:
\begin{align}
    \hat{\Psib},\hat{\Wa},\hat{\Fa},\hat{\Phib} = & \argmin\limits_{\Psib,\Wa,\Fa,\Phib} ||\Y - \Psib \otimes_1 \Wa \otimes_2 \Fa \otimes_3 \Phib||_F^2 \nonumber\\ & + \lambda
    || \text{vec}(\Psib) ||_0, \label{eq:chan_est}
\end{align}
where $||\cdot||_0$ is the $l_0$-norm (number of non-zero terms), $\lambda$ is a scaling parameter, and $\Psib,\Wa,\Fa,\Phib$ are functions of $h_n, \thetarn, \thetatn$, and $d_n$ respectively. The first term in~\eqref{eq:chan_est} minimizes the Frobenius norm of the error and the second term enforces low rank in the estimated tensor reconstruction.  The rank of the reconstructed tensor is the estimate for the number of paths and the second term ensures that a minimal number of paths are used for channel parameter estimation.  Section~\ref{sec:MSVD} provides a method that uses the MSVD to estimate the number of significant paths and recover each of the channel parameters as well as the path gain in \eqref{eq:Tuckermeas}.

\section{Multilinear SVD for mmWave Channel Parameter Estimation}\label{sec:MSVD}
The MSVD is an extension of the singular value decomposition to tensors and reconstructs a third order tensor into a set of column, row, and fiber basis vectors.  This section first introduces the MSVD.  Then, the MSVD of the third order measurement tensor is used to create a reduced rank Tucker form and estimate the channel parameters.  The properties of the MSVD are only minimally covered here and further details can be found in \cite{sidiropoulos2017tensor}.

\subsection{The Multilinear Singular Value Decomposition}
The MSVD reconstructs tensors into a Tucker form, so that the interaction energy between basis vectors is arranged in decreasing order.  The MSVD of the received measurement tensor gives the following Tucker form:
\begin{align}\label{eq:MSVDmeas}
    \begin{split}
        \Y &= \Sigmab \circledcirc_1 \U^{(1)} \circledcirc_2 \U^{(2)} \circledcirc_3 \U^{(3)}, \\
        &= \sum\limits_{i_1=1}^{L_{\rm rx}} \sum\limits_{i_2=1}^{T} \sum\limits_{i_3=1}^{N_{\rm s}} \Sigmab(i_1,i_2,i_3) ~ \uvec{i_1}{(1)} \circledcirc \uvec{i_2}{(2)} \circledcirc \uvec{i_3}{(3)},
    \end{split}
\end{align}
where $\uvec{i_1}{(1)}=\U^{(1)}(:,i_1), \uvec{i_2}{(2)}=\U^{(2)}(:,i_2)$, and $\uvec{i_3}{(3)}=\U^{(3)}(:,i_3)$ are orthonormal basis column vectors such that $(\U^{(i)})^H\U^{(i)}=\bm{I}$ for $i=1,2,3$, where $\bm{I}$ is the identity matrix.  Each basis vector matrix $\U^{(i)}$ is square.  The MSVD is arranged so that a majority of the energy is in the upper left corner of the core tensor $\Sigmab$, corresponding with the strongest interactions between sets of column, row, and fiber vectors.

Singular values for the MSVD are defined such that each dimension of the core tensor $\Sigmab$ has its own set of singular values.  The $l^{\text{th}}$ singular value along the first dimension (or column space) is defined as $||\Sigmab(l,:,:)||_F$, which is the Frobenius norm of the slab of $\Sigmab$ that contains the $l^\text{th}$ column.  The MSVD arranges the singular values so that they decrease as $l$ increases.  Singular values along other dimensions follow similarly.  

It is noted that in the standard SVD, the column space and row space have the same rank.  However, in the multilinear SVD, the column, row, and fiber spaces can have different ranks.  For \eqref{eq:MSVDmeas}, the column rank is the number of columns in $\U^{(1)}$, the row rank is the number of columns in $U^{(2)}$, and the fiber rank is the number of columns in $U^{(3)}$.

\subsection{Rank Reduction}
It is known that few significant paths exist in the model from \eqref{eq:Tuckermeas}, which besides small noise interactions, makes $\Y$ a low rank tensor.  On the other hand, the MSVD from \eqref{eq:MSVDmeas} has a column rank of $L_{\rm rx}$ (number of data streams), a row rank of $\T$ (number of training symbols), and a fiber rank of $N_{\rm s}$ (number of subcarriers).  A majority of the interactions from the MSVD must be eliminated to obtain a low rank representation of $\Y$.  The strongest interactions mainly consist of energy from significant received paths and the weakest interactions are mainly composed of energy from non-significant paths and noise. 

The rank of the MSVD \eqref{eq:MSVDmeas} can be reduced by removing interactions that correspond with singular values below a threshold in each dimension.  This is achieved by removing planes from $\Sigmab_s$ corresponding with the weak singular values along with the interacting row, column, and fiber vectors.  This acts as a denoising process \cite{hayes2017low} and the remaining reduced rank Tucker form is
\begin{equation}\label{eq:MSVDreduced}
    \Y = \Sigmab_s \otimes_1 \U_s^{(1)} \otimes_2 \U_s^{(2)} \otimes_3 \U_s^{(3)},
\end{equation}
where if $r_1,r_2$, and $r_3$ are the reduced ranks of the column, row, and fiber subspaces; then $\Sigmab_s$ is a $r_1 \times r_2 \times r_3$ core tensor, $\U_s^{(1)}$ is a $L_{\rm rx} \times r_1$ matrix, $\U_s^{(2)}$ is a $\T \times r_2$ matrix, and $\U_s^{(3)}$ is a $N_{\rm s}\times r_3$ matrix.  

\begin{figure}[t]
	\includegraphics[width=0.45\textwidth]{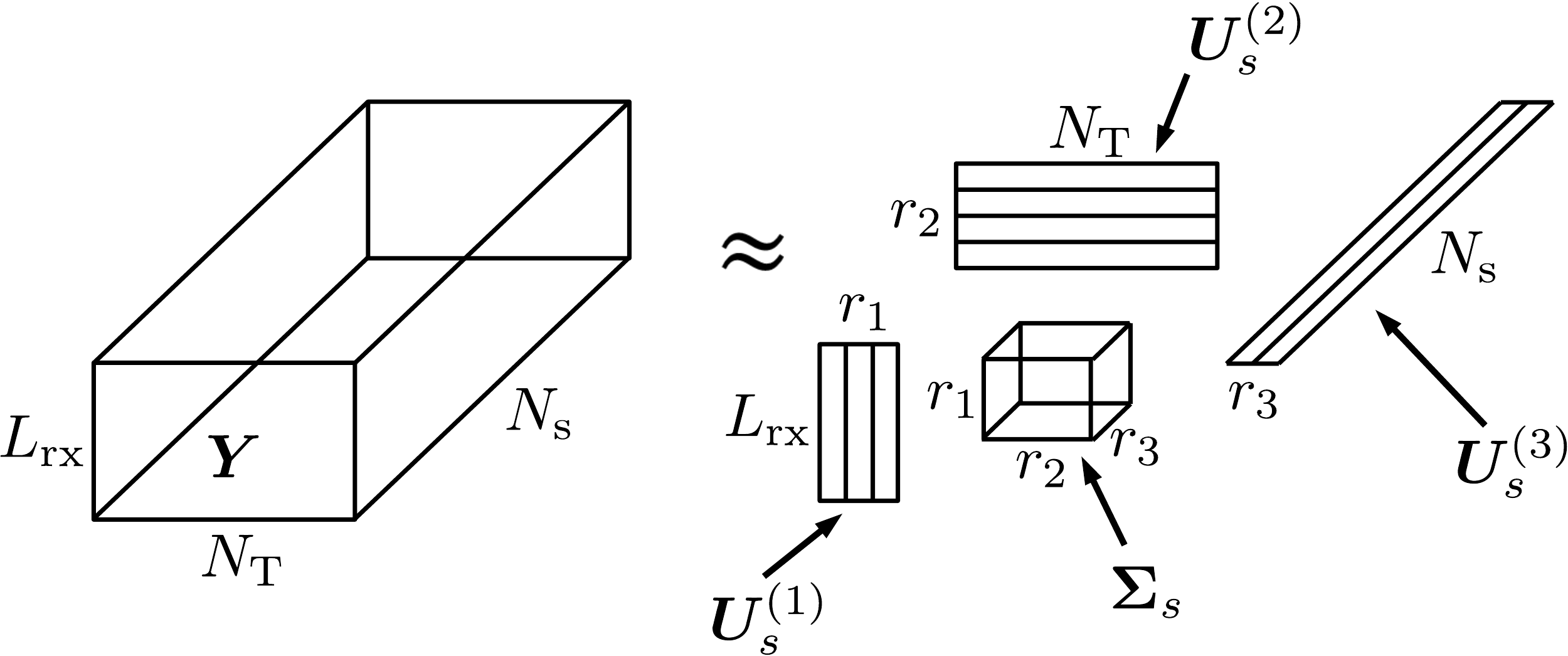}
	\centering
	\caption{The measurement tensor $\Y$ is size $L_{\rm rx} \times \T \times N_{\rm s}$ where $L_{\rm rx}$ is the number of data streams, $\T$ is the number of training symbols, and $N_{\rm s}$ is the number of subcarriers.  The reduced rank Tucker form from the MSVD selects the strongest components from each subspace.}
	\label{fig:SVDblock}
\end{figure}

A variety of methods can be used to select singular value thresholds.  The aim is to select thresholds such that the energy related to weak paths and noise are removed while the remaining terms in \eqref{eq:MSVDreduced} consist of energy from significant paths.  The number of nonzero terms in $\Sigmab_s$ is an estimate for the number of paths $N_{\rm p}$.
The work in \cite{gavish2014optimal} shows that under similar conditions, the optimum threshold for each dimension is $s_\text{thresh}=2.858 s_\text{med}$, where $s_\text{med}$ is the median singular value along that dimension. We use this threshold value, but \cite{gavish2014optimal} discusses alternative thresholds.  Fig.~\ref{fig:SVDblock} shows the reduced rank Tucker tensor form after thresholding.

\subsection{Separating Channel Parameter Estimation into Separate Subspace Problems}
The reduced Tucker form from the MSVD in \eqref{eq:MSVDreduced} eliminates noise and estimates the number of paths $N_{\rm p}$, which is the rank of the measurement tensor.  The denoised and rank reduced MSVD is then approximately equivalent to the noiseless signal in the receiver measurement \eqref{eq:measTuckerFinal} :
\begin{equation}\label{eq:subspaces}
    \Psib \circledcirc_1 \Wa \circledcirc_2 \Fa \circledcirc_3 \Phib
    \approx
    \Sigmab_s \circledcirc_1 \U_s^{(1)} \circledcirc_2 \U_s^{(2)} \circledcirc_3 \U_s^{(3)}.
\end{equation}
Channel parameter estimation is accomplished by estimating the unknown $\Psib$, $\Wa, \Fa,$ and $\Phib$ matrices.  Assuming the reduced Tucker form from the MSVD sufficiently separates the noise from the signal, the basis vectors in \eqref{eq:subspaces} $\U_s^{(1)}, \U_s^{(2)}$, and $\U_s^{(3)}$ share the same subspaces as $\Wa, \Fa,$ and $\Phib$, respectively.  Furthermore, each of the subspaces are independent.  Therefore, representing the channel in Tucker form enables channel parameter estimation to be separated into three independent sub-problems in the column, row, and fiber subspaces, which separately solve for $\thetarn, \thetatn$, and $d_n$, respectively. This significantly reduces the required computation, since otherwise every permutation of $\thetarn, \thetatn$, and $d_n$ has to be considered jointly.

For the column subspace, the objective is to utilize the basis vectors from $\U_s^{(1)}$ to estimate $\Wa$ such that the columns of $\Wa$ correspond with physically realizable paths in the channel.  This equates to finding a value of $\thetarn$ for each column of $\Wa$.  However, this is challenging since the basis vectors generated by the MSVD are not unique. Additionally, the basis vectors in $\U_s^{(1)}$ are orthonormal while the true path column vectors in $\Wa$ are not necessarily orthogonal.  

\begin{figure}[t]
	\includegraphics[width=0.20\textwidth]{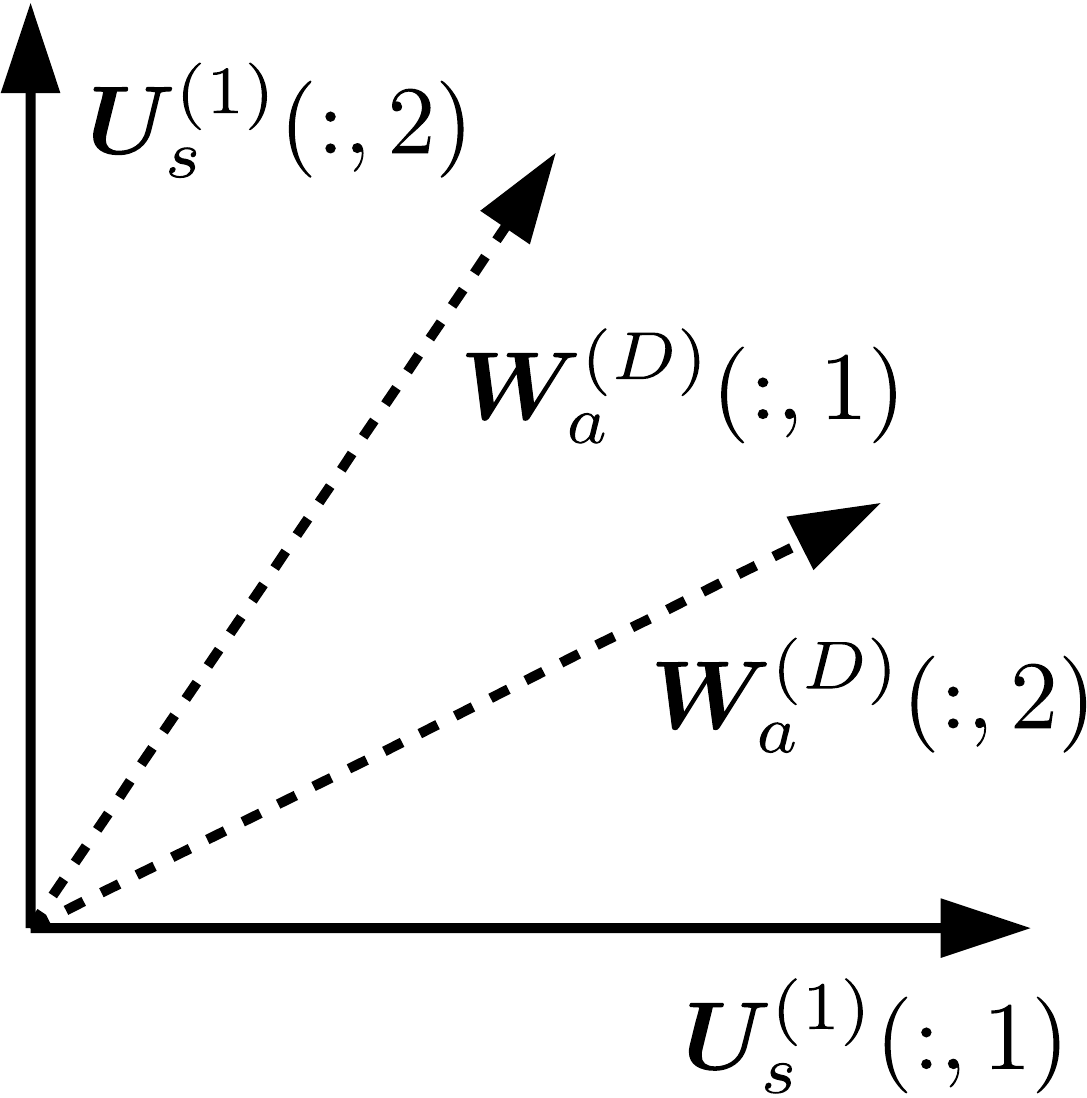}
	\centering
	\caption{A basis transformation converts from an orthogonal MSVD basis to a non-orthogonal dictionary basis.}
	\label{fig:MSVDalign}
\end{figure}

\begin{figure}[t]
	\includegraphics[width=0.40\textwidth]{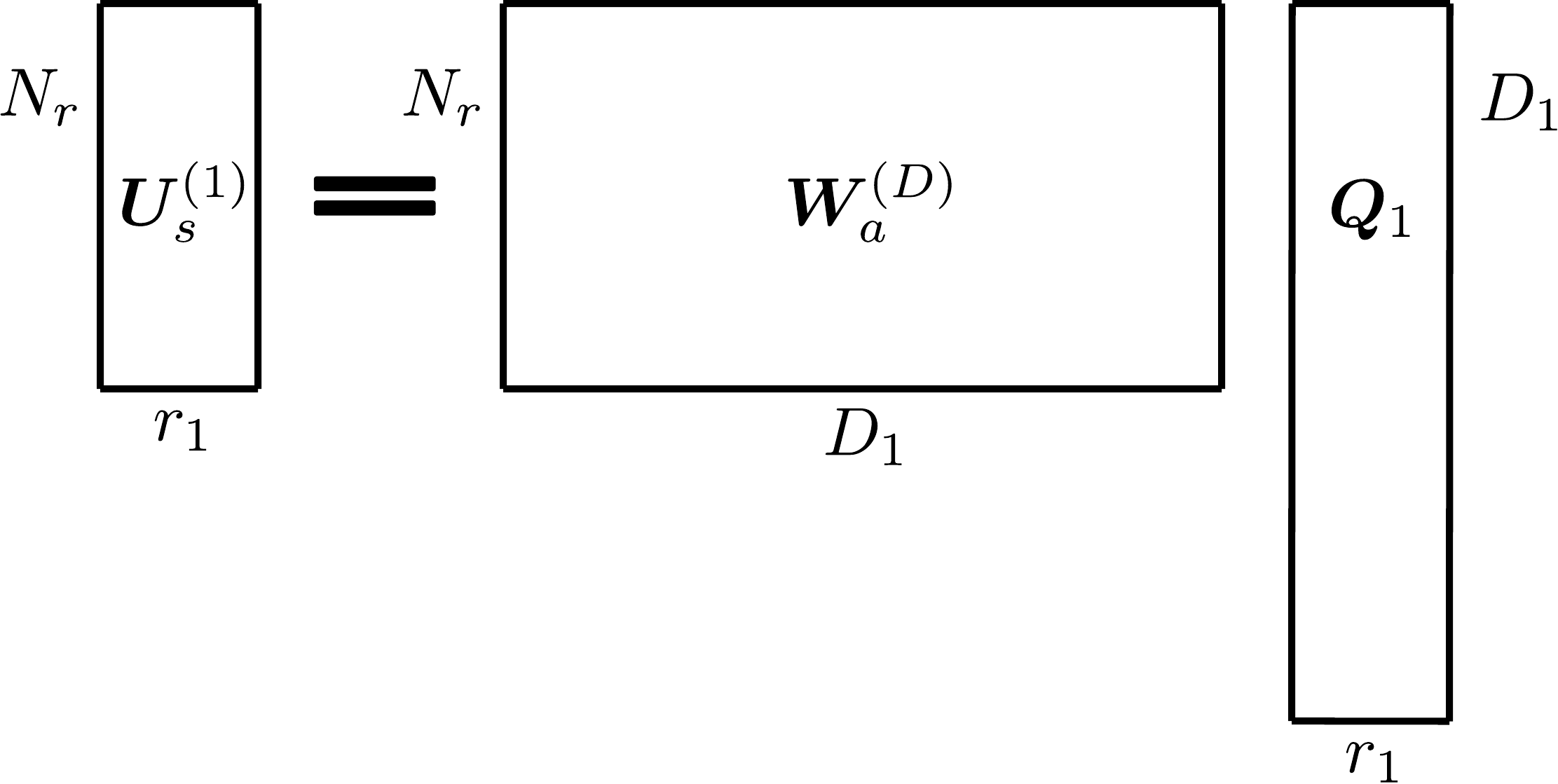}
	\centering
	\caption{Overdetermined problem to compute the column subspace basis transformation $\Q_1$.}
	\label{fig:Q1}
\end{figure}

The desired vectors in the matrix $\Wa$ are unknown, but a dictionary of possible vectors formulated from physical paths exists as follows:
\begin{equation}\label{eq:WaD}
    \WaD = 
    \begin{bmatrix}
    \wanhat(\theta_{\text{rx},1}^{(D)}) & \cdots & \wan(\theta_{\text{rx},D_1}^{(D)})
    \end{bmatrix},
\end{equation}
where $\{\theta_{\text{rx},m}^{(D)}\}_{m=1}^{D_1}$ is a set of $D_1$ dictionary values and
\begin{eqnarray*}
    \wanhat(\theta_{\text{rx},i}) = \frac{\wan(\theta_{\text{rx},i}^{(D)})}{\wan(\theta_{\text{rx},i}^{(D)})^H\wan(\theta_{\text{rx},i}^{(D)})}, & \text{for}~ i=1,\dots,D_1
\end{eqnarray*}
are normalized column vectors from \eqref{eq:wan}, formulated from an over-complete dictionary of $D_1$ possible $\thetarn$ values. Fig.~\ref{fig:MSVDalign} shows an example with two basis vectors and two dictionary vectors.  In this example, the orthonormal basis vectors from the MSVD $\U_s^{(1)}$ do not align with possible basis vectors in $\WaD$.  Note that Fig.~\ref{fig:MSVDalign} is a projection into two dimensions while the actual basis vectors are in $N_{\rm rx}$ space.  

Estimating the columns of $\Wa$ essentially becomes a basis transformation where we seek the basis transformation matrix $Q_1$ such that:
\begin{equation}\label{eq:Q1unsolved}
    \U_s^{(1)} = \WaD\Q_1.
\end{equation}
This is an overdetermined problem since the dictionary is overcomplete and there are many more dictionary terms than paths ($D_1 \geq N_{\rm p}$).  Thus, the solution is not guaranteed to be unique. However, the measurement tensor is low rank and the solution for $Q_1$ must have $r_1$ nonzero terms to match the rank of the column subspace. This leads to a sparse optimization problem, where the selection of dictionary basis vectors is accomplished by enforcing sparsity on $\Q_1$.  The row and column subspaces follow similarly.

\subsection{Subspace Estimation}
Similar sparse estimation problems to \eqref{eq:Q1unsolved} are posed in \cite{hayes2017low}, \cite{cotter2005sparse}, \cite{malioutov2005sparse} where sparsity is an outcome of vector selection from an overcomplete dictionary.  In these works sparsity is enforced in $\Q_1$ by solving the optimization problem:
\begin{equation}
    \arg \min\limits_{\Q_1} ||\U_s^{(1)}-\WaD \Q_1||^2_F + \lambda||\Q_1^{l_2}||_1,
\end{equation}
where $\lambda$ is a tuning parameter and $\Q_1^{l_2}$ is a vector containing the $l_2$ norm for each row of $\Q_1$.  Posing the sub-problem in this form results in the multiple measurement vector sparse estimation problem. Multiple solution techniques are discussed in \cite{cotter2005sparse}, \cite{chen2005sparse}, but we choose to use the multiple measurement vector orthogonal matching pursuit (M-OMP) algorithm.  M-OMP is chosen since it is a greedy algorithm and requires less computation.  Details on M-OMP can be found in \cite{chen2005sparse}.  

The sparsity of the column subspace $\Q_1$ is set to the rank of the column subspace found from the MSVD ($r_1$), which is the number of columns in $\U_s^{(1)}$.  The output of M-OMP is $\Q_1$ with $r_1$ non-zero rows.  This effectively eliminates all but $r_1$ dictionary vectors and allows the dictionary to be reduced to the following: 
\begin{equation}
    \WaDTilde = \WaD \bm{A},
\end{equation}
where the reduced dictionary matrix $\WaDTilde$ is $N_{\rm rx} \times r_1$ and $\bm{A}$ is a sparse matrix identical to $Q_1$, but with any non-zero rows replaced with a row of ones.  Then the reduced dictionary is used in replacement of \eqref{eq:Q1unsolved} to obtain
\begin{equation}\label{eq:Q_1t}
    \U_s^{(1)} = \WaDTilde \Tilde{\Q_1},
\end{equation}
where $\Tilde{\Q_1}$ is a $r_1 \times r_1$ matrix.  Note that selecting $r_1$ dictionary terms at this step significantly reduces future computational effort.  Similarly, the row subspace uses a reduced $r_2 \times r_2$ matrix $\Tilde{\Q}_2$ and a $\T \times r_2$ matrix $\FaDTilde$ such that
\begin{equation}\label{eq:Q_2t}
    \U_s^{(2)} = \FaDTilde \Tilde{\Q}_2.
\end{equation}
The fiber subspace uses a reduced $r_3 \times r_3$ matrix $\Tilde{\Q}_3$ and a $N_{\rm s} \times r_3$ matrix $\PhibDTilde$ such that
\begin{equation}\label{eq:Q_3t}
    \U_s^{(3)} = \PhibDTilde \Tilde{\Q}_3.
\end{equation}
At this point dictionary terms have been selected in each subspace.  The transformation matrices $\Tilde{\Q}_i$ for $i=1,2,3$ contain information about how each of the dictionary vectors align with the MSVD basis vectors, but dictionary vectors have not yet been explicitly chosen as estimates for the basis vectors.

\subsection{Super-Resolution Channel Parameter Estimation}
The dictionaries used to obtain solutions to \eqref{eq:Q_1t}-\eqref{eq:Q_3t} may be coarse since the dictionaries are limited in size.  This is especially true since the computational effort of M-OMP increases with the number of dictionary terms and smaller coarse dictionaries reduce computations.  Higher resolution can be obtained by iteratively updating the dictionary as strong dictionary terms are identified.  One approach for this is the K-SVD method, which is a dictionary learning algorithm that can be employed with sparse problems \cite{aharon2006k,rubinstein2008efficient}.  We use K-SVD for our simulations as it significantly reduced computation time when compared to a single large dictionary.  Each iteration begins by solving the sparse estimation problem with the last dictionary set.  Then a new dictionary is created that focuses around the dictionary terms chosen during the sparse step.

\subsection{MSVD Basis Transformation}

The best dictionary terms have been selected from \eqref{eq:Q_1t}-\eqref{eq:Q_3t} along with their transformation matrices.  These are now used to transform the MSVD in terms of the dictionary basis set in every subspace.  Substituting $\U_s^{(1)}$, $\U_s^{(2)}$, and $\U_s^{(3)}$ from \eqref{eq:Q_1t}, \eqref{eq:Q_2t}, and \eqref{eq:Q_3t} into \eqref{eq:MSVDreduced}:
\begin{equation}
    \Y = \Sigmab_s \circledcirc_1 \big( \WaDTilde \Tilde{\Q}_1 \big) \circledcirc_2 \big( \FaDTilde \Tilde{\Q}_2 \big) \circledcirc_3 \big( \PhibDTilde \Tilde{\Q}_3 \big).
\end{equation}
Or, equivalently:
\begin{equation}\label{eq:Ytrans}
    \Y = \Sigmab_s' \circledcirc_1  \WaDTilde  \circledcirc_2  \FaDTilde  \circledcirc_3  \PhibDTilde,
\end{equation}
where 
\begin{equation}
    \Sigmab_s' = \Sigmab_s \circledcirc_1 \Tilde{\Q}_1 \circledcirc_2 \Tilde{\Q}_2 \circledcirc_3 \Tilde{\Q}_3.
\end{equation}
The tensor in \eqref{eq:Ytrans} expresses $\Y$ in terms of the basis created by the dictionaries in each subspace.  The core tensor $\Sigmab_s'$ contains the interaction energy between the path channel parameters $\thetarn$, $\thetatn$, and $d_n$ from the dictionary.  However, it may not be obvious which path channel parameters have the strongest interactions and should be linked together.

\subsection{Linking Channel Parameters to Paths}
It is known that the tensor $\Y$ has rank $N_{\rm p}$.  This low rank structure provides a means to select and link the dictionary terms to paths.  One may expect $\Sigmab_s'$ in \eqref{eq:Ytrans} to have $N_{\rm p}$ non-zero terms.  However, this is not guaranteed after the basis transformation as the dictionary vectors in each subspace are generally not orthogonal.  Therefore, we must eliminate interactions from $\Y$ so that it again has rank $N_{\rm p}$.  To do this, we select the dictionary terms with the strongest interactions in $\Sigmab_s'$.  This is accomplished by first taking the MSVD of $\Sigmab_s'$:
\begin{equation}\label{eq:doubleMSVD}
    \Sigmab_s' = \Sb \circledcirc_1 \Db^{(1)} \circledcirc_2 \Db^{(2)} \circledcirc_3 \Db^{(3)}.
\end{equation}
The MSVD of $\Sigmab_s'$ results in the core tensor $\Sb$, which organizes the interaction energies in decreasing order.

At this point the strongest $N_{\rm p}$ elements in $\Sb$ correspond with the $N_{\rm p}$ paths.  It is noted that distinct channel parameters for each path lead to $N_{\rm p} = r_1 = r_2 = r_3$ and the strongest interactions will be the diagional elements in $\Sb$.  However, there are scenarios when this is not the case.  For example, if $N_{\rm p} = 2$, but both paths have the same $\thetarn$ with different $\thetatn$ and $d_n$ then $r_1 = 1$ while $r_2 = 2$ and $r_3 = 2$.  In this case, one of the strongest interactions will be off-diagonal in $\Sb$.  The method in \cite{zhou2017low} does not offer a solution for this type of scenario since the CPD tensor form does not allow multiple paths to share parameter vectors.

\begin{figure}[t]
	\includegraphics[width=0.49\textwidth]{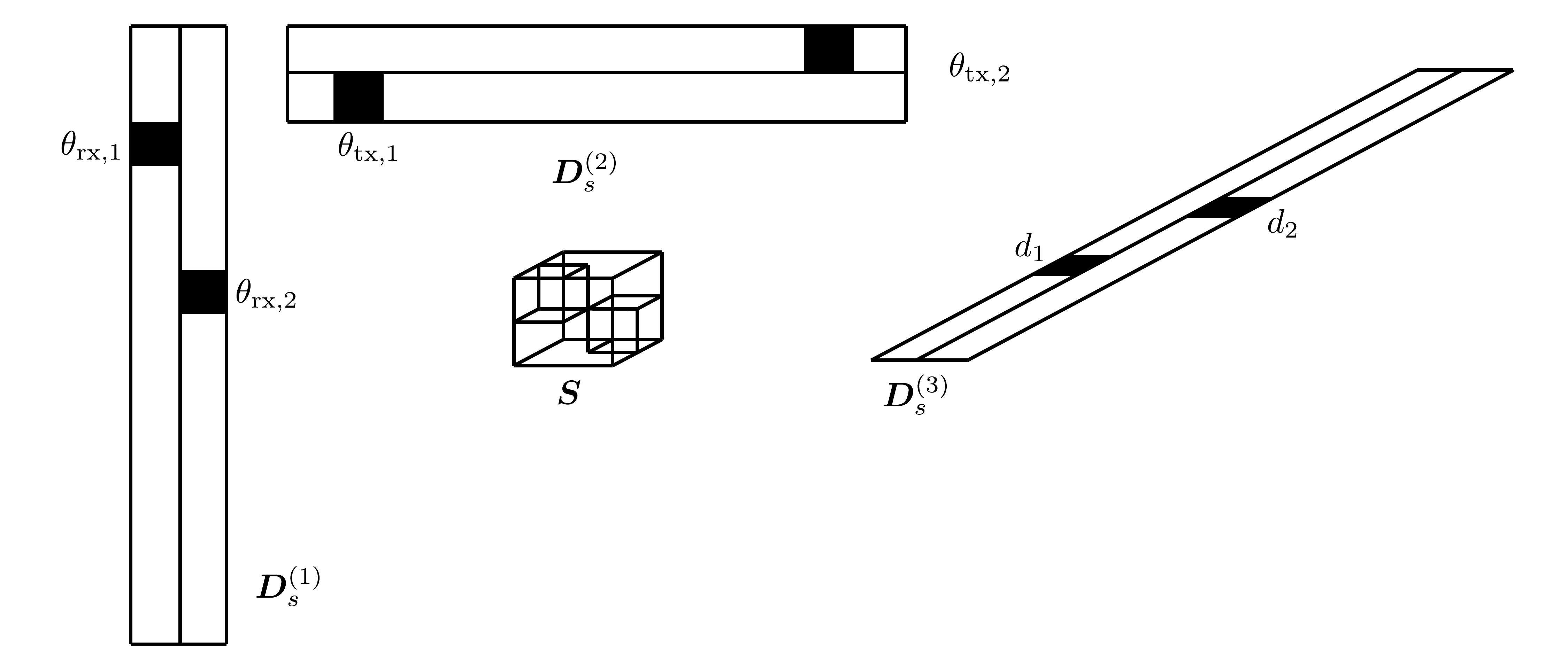}
	\centering
	\caption{Linking channel parameters for each path.}
	\label{fig:linkPath}
\end{figure}

A significant advantage to the MSVD in \eqref{eq:doubleMSVD} is that it provides a simple means to determine the values of $\thetarn$, $\thetatn$, and $d_n$ for each path.  In the form of \eqref{eq:doubleMSVD} each column of $\Db^{(i)}$ for $i=1,2,3$ corresponds with a path and has elements that correspond with the amount that each dictionary term is aligned for that subspace and path.  Therefore, the channel parameters for each path are estimated by choosing the maximum value in each column of $\Db^{(i)}$ for $i=1,2,3$.  Suppose that the strongest $N_{\rm p}$ paths in $\Sb$ have been chosen and the channel parameters are desired for path $n$ that correspond with the element $\Sb(i_1,i_2,i_3)$. Then, the channel parameters are found as:
\begin{align}\label{eq:channelParam_m}
    \begin{split}
        m_1 &= \arg \max\limits_{m} \Db(m,i_1), \\
        m_2 &= \arg \max\limits_{m} \Db(m,i_2), \\
        m_3 &= \arg \max\limits_{m} \Db(m,i_3).
    \end{split}
\end{align}
As a result, the channel parameters for that path are:
\begin{align}\label{eq:ChannParamEst}
        \theta_{\text{rx},n} = \Tilde{\theta}_{\text{rx},m_1}^{(D)}, \quad 
        \theta_{\text{tx},n} = \Tilde{\theta}_{\text{tx},m_2}^{(D)}, \quad 
        d_n = \Tilde{d}_{m_3}^{(D)}~,
\end{align}
where $\{\Tilde{\theta}_\text{rx}^{(D)}\}$, $\{\Tilde{\theta}_\text{tx}^{(D)}\}$, $\{\Tilde{d}^{(D)}\}$ are the set of dictionary terms from \eqref{eq:Q_1t}-\eqref{eq:Q_3t} with sizes $r_1$, $r_2$, and $r_3$, respectively.  
An example of the path linking process is seen in Fig.~\ref{fig:linkPath}.  In this example, the strongest two paths are the diagonal elements of $\Sb_s$.  The strongest elements in each column are selected and used to link and estimate the channel parameters.

\subsection{Estimating Path Gain}
If desired, the path gain can be estimated for each path using the measurement vector $\Y$ and the estimated channel parameters.  This is done by vectorizing the measurement signal tensor as shown in \cite{sidiropoulos2017tensor}:
\begin{equation}
    \vect{\Psib \circledcirc_1 \Wa \circledcirc_2 \Fa \circledcirc_3 \Phib} = (\Phib \odot \Fa \odot \Wa) \bm{h},
\end{equation}
where $\odot$ is the Khatri-Rao product and $\bm{h}$ is a vector of the path gains.  Let $\bm{A} = \Phib \odot \Fa \odot \Wa$;  then, the path gains are  estimated by solving for $\bm{h}$ such that 
\begin{equation}
    \arg \min\limits_{\bm{h}} \Big|\Big| \Y - \bm{A} \bm{h} \Big|\Big|^2.
\end{equation}
This is solved using the pseudoinverse or SVD methods \cite{hayes2017low}.

\begin{algorithm}[t]
\caption{Channel Parameter Estimation: $\{\thetarn,\thetatn,d_n,h_n\}_{n=1}^{N_{\rm p}} = $estMSVD$(\Y)$}
\label{alg:1}
\begin{algorithmic}\label{alg:channelParamEst}
\STATE $1)$ Take the MSVD of the measurement tensor:\\
~~~~$\Y = \Sigmab \circledcirc_1 \U^{(1)} \circledcirc_2 \U^{(2)} \circledcirc_3 \U^{(3)}$
\STATE $2)$ Threshold the singular values and estimate $N_{\rm p}$:\\
~~~~$\Y \approx \Sigmab_s \circledcirc_1 \U_s^{(1)} \circledcirc_2 \U_s^{(2)} \circledcirc_3 \U_s^{(3)}$
\STATE $3)$ Use K-SVD and M-OMP to solve for the dictionary\\ ~~~~vectors ($\WaDTilde,\FaDTilde,\PhibDTilde$) and trasformation matrices\\ ~~~~$\Tilde{\Q}_1,\Tilde{\Q_2},\Tilde{\Q_3}$.\\
\STATE $4)$ Convert the MSVD to dictionary basis:\\
~~~~$\Y = \Sigmab_s' \circledcirc_1  \WaDTilde  \circledcirc_2  \FaDTilde  \circledcirc_3  \PhibDTilde$
\STATE $5)$ Take the MSVD of the core tensor:\\
~~~~$\Sigmab_s' = \Sb \circledcirc_1 \Db^{(1)} \circledcirc_2 \Db^{(2)} \circ_3 \Db^{(3)}$
\STATE $6)$ Obtain $(\thetarn,\thetatn,d_n)$ for $n=1,\dots,N_{\rm p}$ by selecting\\ ~~~~the strongest elements in $\Sb$ along with the maximum\\ ~~~~elements in the corresponding columns.\\
\STATE $7)$ Estimate $h_n$ using $\Y$ and the estimated channel\\ ~~~~parameters.
\end{algorithmic}
\end{algorithm} 

\subsection{Applications of Channel Parameter Estimation}

Algorithm \ref{alg:1} summarizes the proposed channel parameter estimation algorithm.  Once obtained, the estimates for the channel parameters can be used to estimate the channel matrix ($\Hmat[k]$) for $k=1,\dots,N_{\rm s}$ by substituting $\thetarn$, $\thetatn$, $\tau_n = d_n/c$ and $h_n$ for the $N_{\rm p}$ paths into \eqref{eq:channel2}.  The channel parameters from one LOS path or three NLOS paths are also sufficient for estimating receiver location and orientation as well as mapping the environment \cite{shahmansoori2018position},\cite{ruble2018}. 

\section{Waveform Considerations for MIMO OFDM Channel Parameter Estimation}\label{sec:wavePar}
Channel parameter estimation accuracy depends on the communication waveform.  Certain waveform parameters improve channel parameter estimation accuracy while other parameters can significantly degrade channel parameter estimation accuracy.  This section provides details on how waveform parameters effect channel parameter estimation as well as the validity of the assumptions used to reach the measurement model in \eqref{eq:measTuckerFinal}.

\subsection{Frequency Selectivity}\label{sec:FreqSel}
It is assumed in \eqref{eq:channel2} that the path gain is constant over all subcarriers, or $h_n(k) \approx h_n$ for all $k$.  This subsection discusses waveform requirements for this frequency non-selective assumption to be valid, which depends on the delay spread of the channel and channel dispersion effects.  

\paragraph{Delay Spread}
Frequency selective fading occurs when the multipath delay extends over more than one symbol~\cite{hampton2013introduction}.  A cyclic prefix in each OFDM symbol aims to capture multipath from previous symbols. However, the frequency non-selective channel assumption requires the OFDM symbol length $T_\text{OFDM}$ to be much larger than the delay spread ($\sigma_\tau$), or $T_\text{OFDM} >> \sigma_\tau$.

\paragraph{Channel Dispersion}
For large OFDM bandwidths and large arrays, small phase differences observed at the antenna elements for each subcarrier can cause channel dispersion in space and time \cite{shahmansoori2018position},\cite{brady2015wideband}.  Channel dispersion increases with the number of antenna elements and 
subcarriers.  The frequency non-selective assumption holds if $N_{\rm rx} N_{\rm s} /2T_\text{OFDM}<f_c$, where channel dispersion effects become more noticeable as the term to the left of the inequality increases.  Larger center frequencies lead to less channel dispersion.  

Whether dealing with delay spread or channel dispersion effects, a frequency non-selective assumption requires the symbol duration $\To$ to be sufficiently large.  On the other hand, larger OFDM bandwidths and additional antenna elements cause channel dispersion.  
Systems with channel dispersion can still achieve frequency non-selectivity by only utilizing a subset of subcarriers and dividing channel parameter estimation into blocks where the subcarriers in each block do not observe channel dispersion.  Alternatively, the channel parameter estimation algorithm can be adjusted to account for the phase differences in the subcarriers in the model.

\subsection{Effective SNR}
Channel parameter estimation improves as the number of subcarriers $N_{\rm s}$ increases and the symbol length of the training sequence $\T$ increases.  Each of these adds dimensionality to the measurement in \eqref{eq:measTuckerFinal} and provides additional observations of the same channel parameters.  The additional diversity of measurements increases the effective SNR and leads to improved 
parameter estimation.  Also, increasing the numbers of antenna elements at the transmitter and receiver increases the measurement tensor dimensions and effective SNR.

\subsection{Distance Redundancy}
A limitation of channel parameter estimation is that the solution for the path distance is not unique.  The path distance is estimated in the tensor fiber subspace.  From \eqref{eq:phin}, the fiber vector for each path contains the terms:
\begin{eqnarray*}
    \phib_n(d_n) = \Big\{ \e^{ \frac{-j 2\pi k d_n}{c ~ \To} } \Big\}_{k=0}^{N_{\rm s}-1} & \text{for}~ n = 0\dots,N_{\rm p}-1.
\end{eqnarray*}
For path $n$, multiple values of $d_n$ give the same $\phib_n(d_n)$. To see this, first let $d_\text{OFDM} = c T_\text{OFDM}$.  Then, we can write: 
\begin{equation}
    \phib_n(d_n+d_\text{OFDM}) = \phib_n(d_n).
\end{equation}
Therefore, the solution to the distance for each path is periodic with period $d_\text{OFDM}$.  A unique solution is only achieved if the distance search space is limited to a range of $d_\text{OFDM}$.  This guarantees that multiple distances are not found for each path.  This effect can be mitigated by increasing $T_\text{OFDM}$, which increases $d_\text{OFDM}$ and allows a larger distance search space with a unique distance solution.

\section{Simulation Results}\label{sec:SimsAndResults}

\begin{table}[t]
    \begin{center}
    \begin{tabular}{|c|c|c|}
    \hline
    \multicolumn{3}{|c|}{\textbf{Waveform Specifications}}\\
    \hline
    Center Frequency & $f_c$ & 60 GHz\\
    \hline
    Number of Subcarriers & $N_{\rm s}$ & 10-100\\
    \hline
    OFDM Bandwidth & $B_\text{OFDM}$ & 100 MHz\\
    \hline
    OFDM Symbol Length & $T_\text{OFDM}$ & 1 $\mu$s\\
    \hline
    Sampling Time & $T_s$ & 10 ns\\
    \hline
    Subcarrier Spacing & $\Delta f_s$ & 1 MHz\\
    \hline
    Training Sequence Length & $\T$ & 10-100 symbols\\
    \hline
    \end{tabular}
    \end{center}
    \caption{Range of waveform specifications considered in simulations.}
    \label{table:specs}
\end{table}

This section compares simulations of the proposed MSVD channel parameter estimation to the CRB bound under a variety of waveform parameters, numbers of antenna elements, and SNR.  The range of waveform parameters considered are listed in Table \ref{table:specs}.  The CRB bound changes for different path geometries.  Thus, we choose a single path geometry for all of our simulations, which consists of one LOS path and one NLOS path between a transmitter and receiver as seen in Fig.~\ref{fig:SNRpaths}. We fix the array orientations to $\phi_{\rm rx}=0$ and $\phi_{\rm tx}=0$. The SNR, number of training symbols, and number of subcarriers are varied while maintaining this geometry.

Simulations are conducted by first calculating the channel parameters $\thetarn$, $\thetatn$, and $d_n$ dictated by the path geometry in Fig.~\ref{fig:SNRpaths}.  The path gain for the first path is set to $h_1 = 1$ and the path gain for the second path is set to $h_2 = 0.5$.  Similar to \cite{suryaprakash2016millimeter}, the precoding and combining matrices $\F[k]$ and $\W[k]$ from \eqref{eq:meas1} are generated by uniformly sampling from $\{1,-1,i,-i\}$ for each element.  Then, the measurement tensor $\Y$ from \eqref{eq:measTuckerFinal} is constructed from the channel parameters, path gain, precoding and mixing matrices, and additive random noise to achieve a specified SNR.  It should be noted that we do not explicitly simulate interfering paths and weaker reflecting paths.  The random noise is used to represent interfering paths or many non-significant paths.   

Simulations are compared to the CRB bound through the root-mean square error (RMSE) for each channel parameter, which for path $n$ is calculated as:
\begin{align}
    \text{RMSE}(\thetarn) &= \sqrt{E[(\thetarn-\thetarn^0)^2]},\\
    \text{RMSE}(\thetatn) &= \sqrt{E[(\thetatn-\thetatn^0)^2]},\\
    \text{RMSE}(d_n) &= \sqrt{E[(d_n-d_n^0)^2]},
\end{align}
where $\thetarn^0$, $\thetatn^0$, and $d_n^0$ are the true channel parameters.  The expectation value is simulated with $N_\text{sim}$ Monte-Carlo simulations such that for the random variable $x$,
\begin{equation}
    E[x^2] = \frac{\sum_{i=1}^{N_\text{sim}}x_i^2}{N_\text{sim}},
\end{equation}
where each $x_i$ is a Monte-Carlo simulated observation of $x$.

The CRB bound provides a RMSE performance bound for any estimator and is calculated using the inverse of the Fisher information matrix $\bm{I}(\thetab)$ \cite{kay1993statistical},\cite{sidiropoulos2017tensor}, which is derived in Appendix \ref{sec:CRB} to give:
\begin{equation}
    \bm{C}(\thetab) = \sqrt{\bm{I}^{-1}(\thetab)},
\end{equation}
where $\bm{C}(\thetab)$ is the CRB bound matrix.  The vector $\thetab$ is the collection of channel parameters and path gains for all paths:
\begin{equation}
    \thetab = 
    \begin{bmatrix}
        \{\thetarn\}_1^{N_{\rm p}}, & \{\thetatn\}_1^{N_{\rm p}}, & \{d_n\}_1^{N_{\rm p}} & \{h\}_1^{N_{\rm p}}
    \end{bmatrix}.
\end{equation}
The diagonal elements of $\bm{C}(\thetab)$ are the RMSE bounds for each of the corresponding elements in $\thetab$.

\subsection{SNR and Numbers of Antenna Elements}

\begin{figure}[t]
	\includegraphics[width=0.45\textwidth]{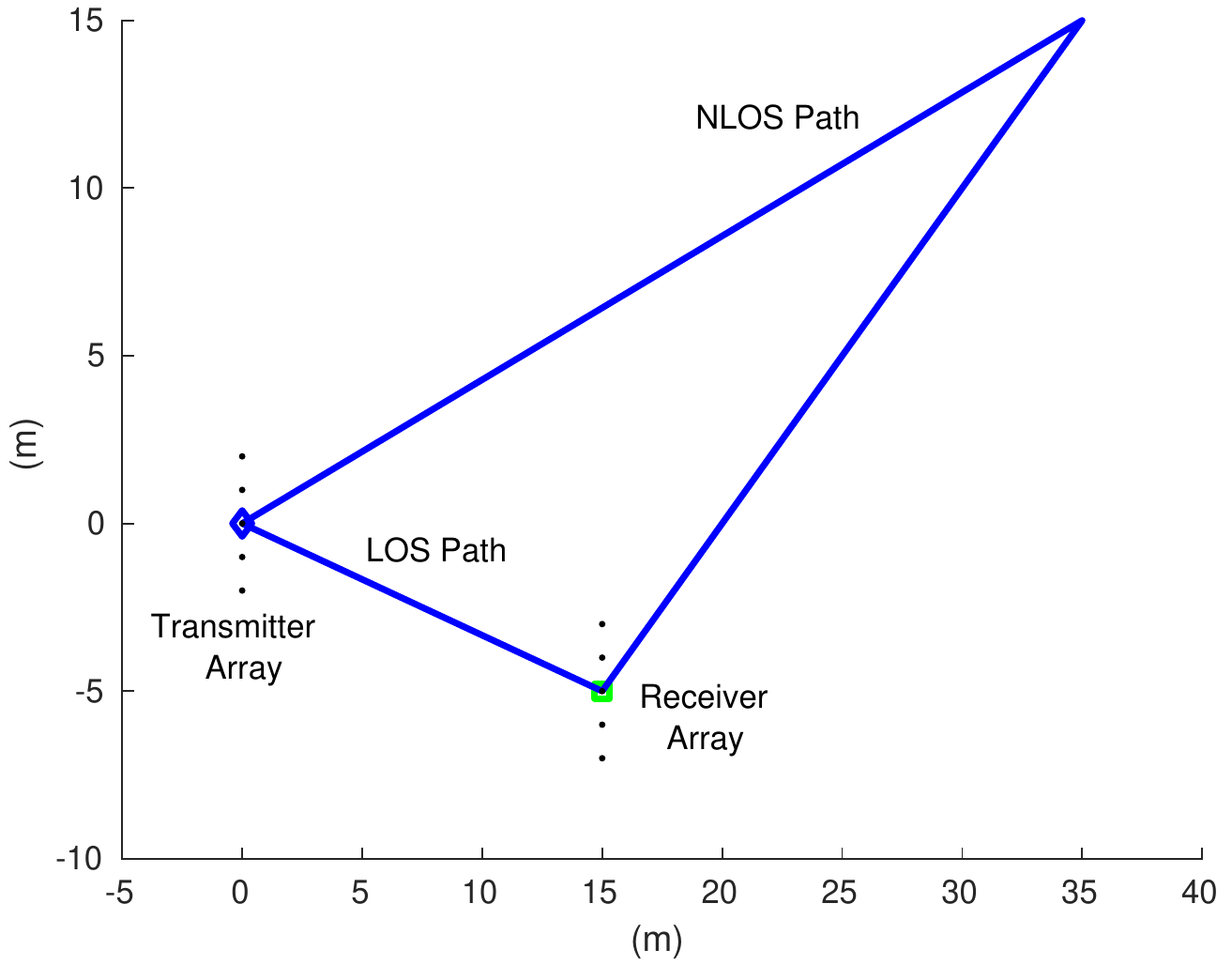}
	\centering
	\caption{Path geometry for channel parameter simulations.}
	\label{fig:SNRpaths}
\end{figure}
     
\begin{figure*}
     \centering
     \begin{subfigure}[b]{0.49\textwidth}
         \centering
         \includegraphics[width=\textwidth]{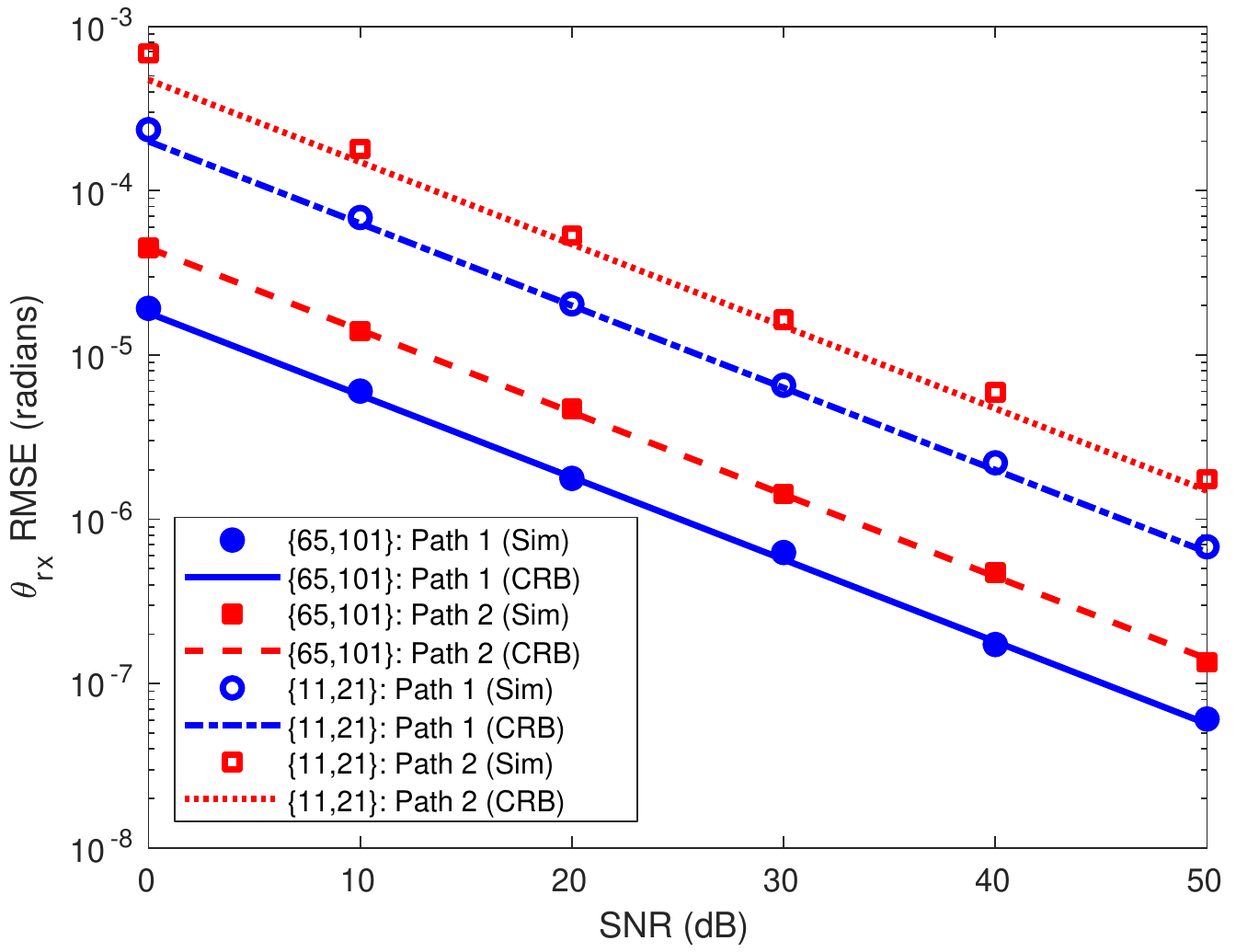}
         \caption{}
         \label{fig:SNRthetar}
     \end{subfigure}
     \hfill
     \begin{subfigure}[b]{0.49\textwidth}
         \centering
         \includegraphics[width=\textwidth]{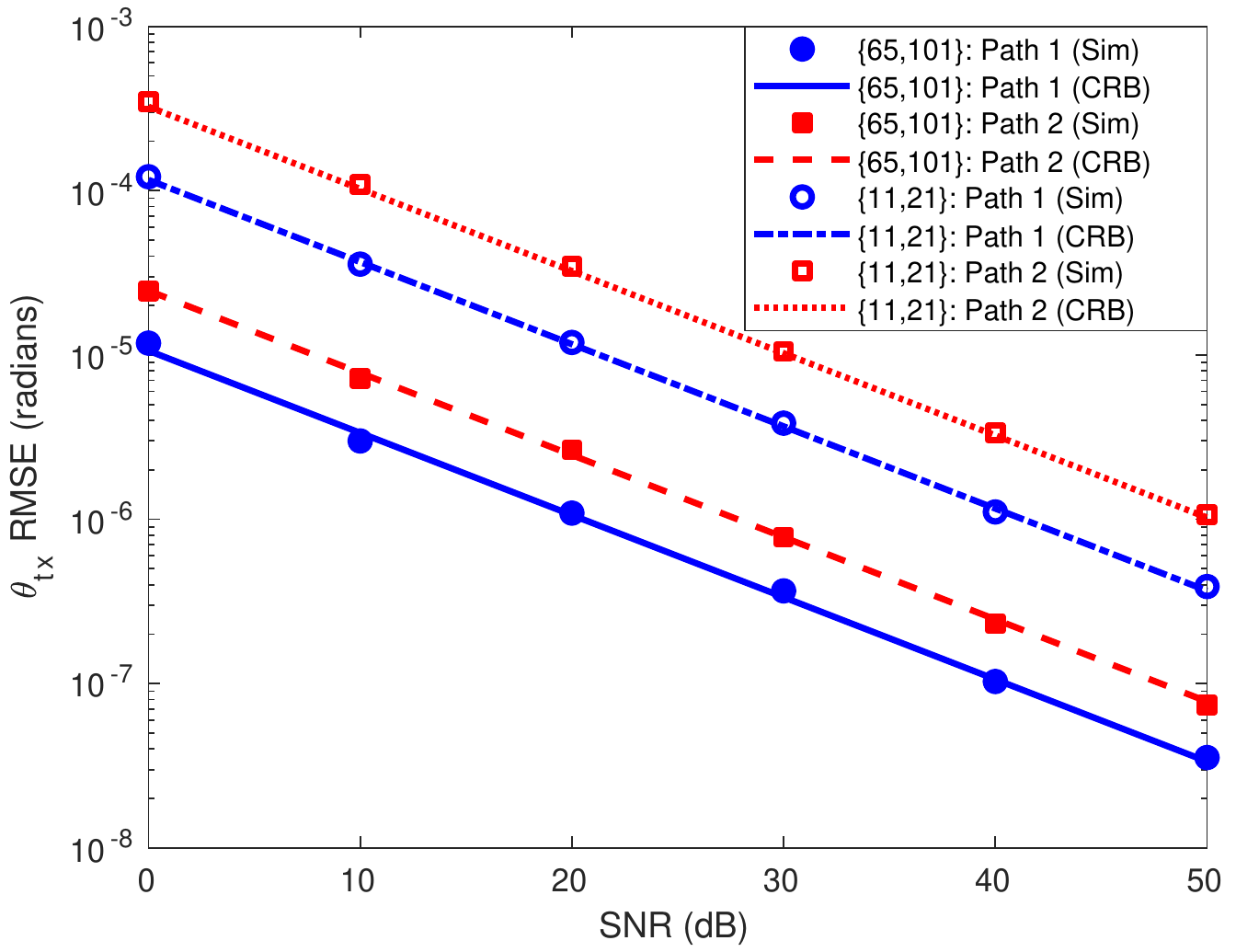}
         \caption{}
         \label{fig:SNRthetat}
     \end{subfigure}
     \\
     \begin{subfigure}[b]{0.49\textwidth}
         \centering
         \includegraphics[width=\textwidth]{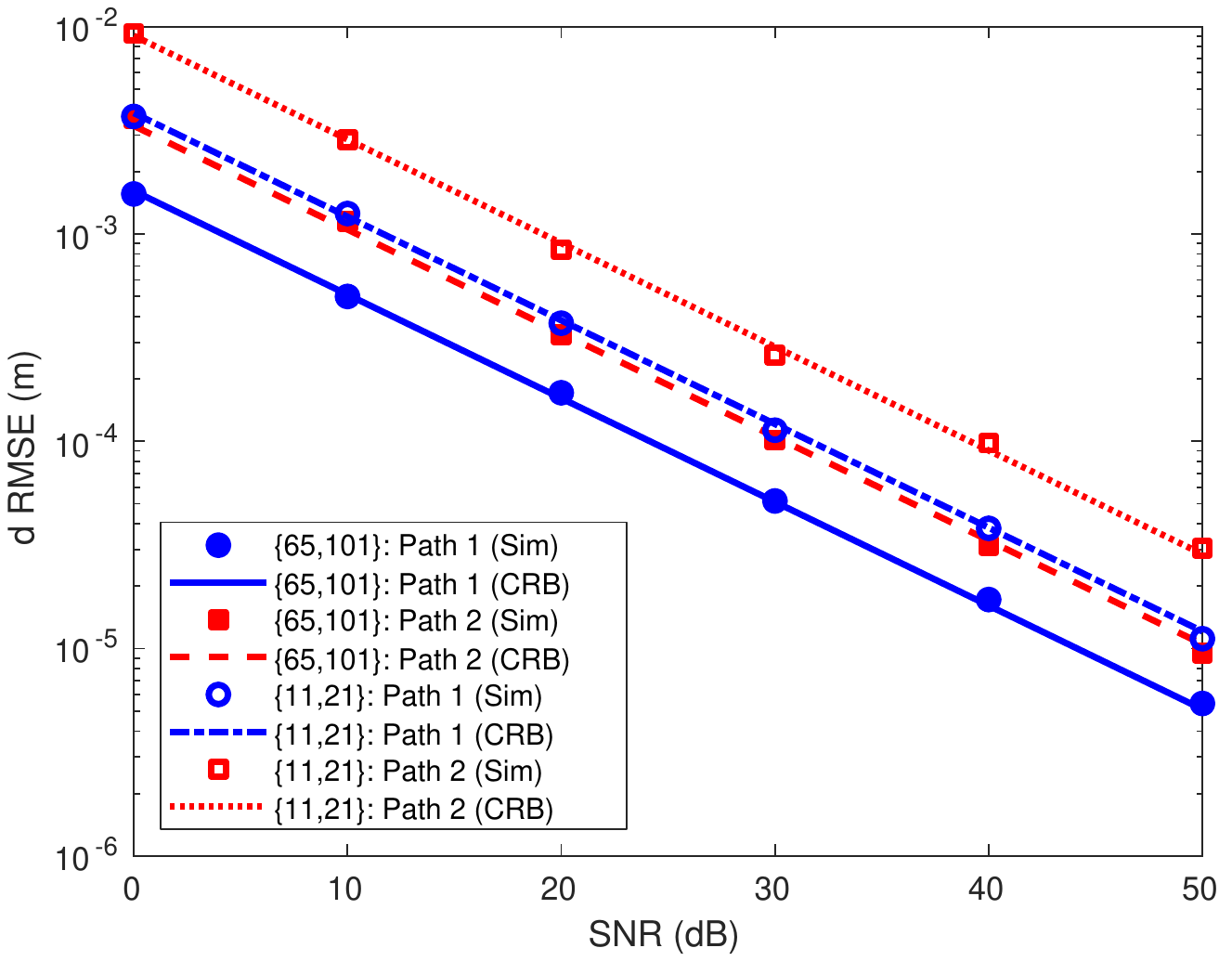}
         \caption{}
         \label{fig:SNRd}
     \end{subfigure}
     \hfill\begin{subfigure}[b]{0.49\textwidth}
         \centering
         \includegraphics[width=\textwidth]{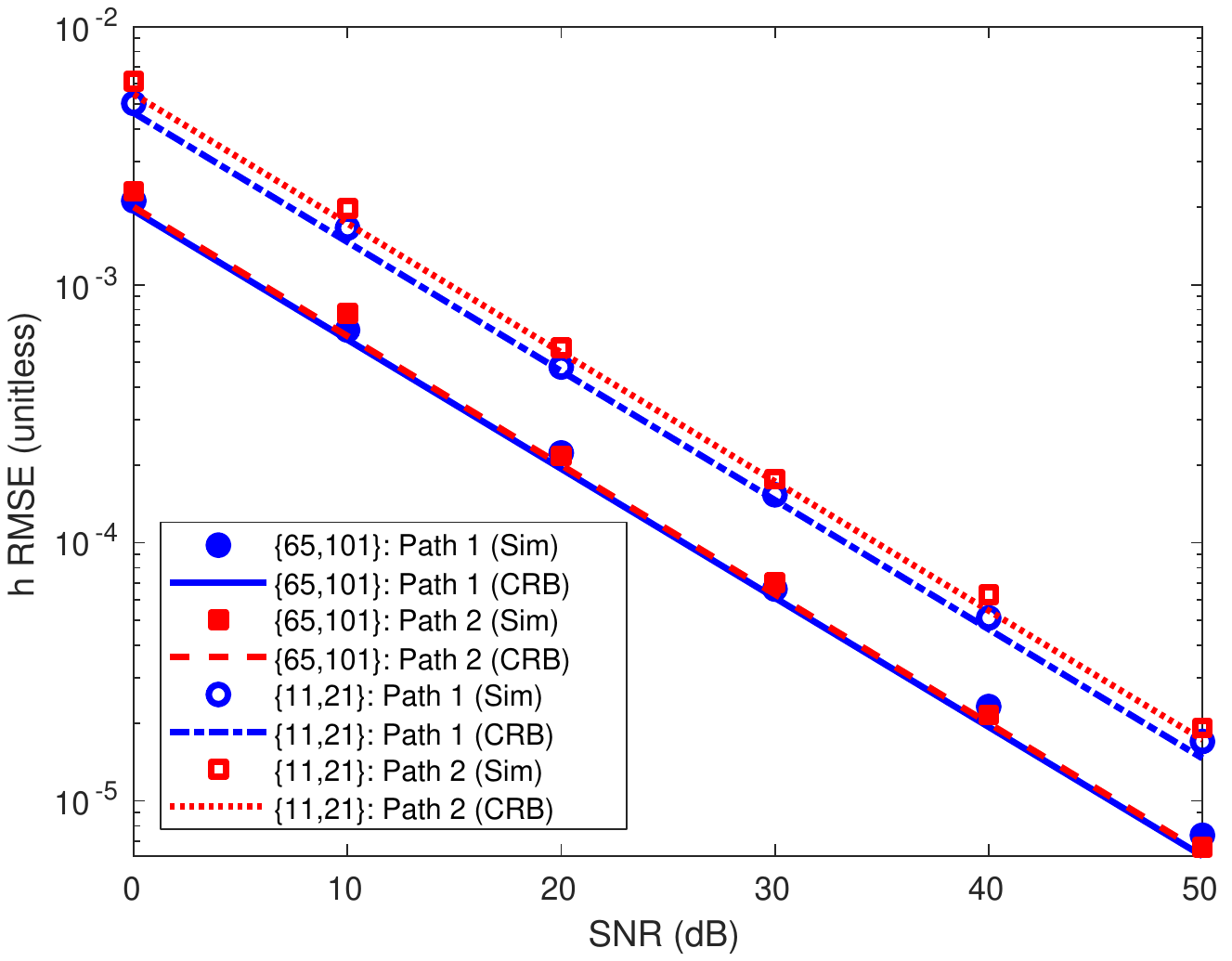}
         \caption{}
         \label{fig:SNRh}
     \end{subfigure}
        \caption{Simulated channel parameter estimation RMSE and CRB for the paths in Fig.~\ref{fig:SNRpaths} for $N_{rx}=11, N_{tx}=21$ and $N_{rx}=65, N_{tx}=101$. Plots (a), (b), (c), and (d) show $\thetarn$, $\thetatn$, $d_n$, and $h_n$ respectively for $n=1,2$.}
        \label{fig:SNR}
\end{figure*}

Channel parameter estimation error is first studied as a function of SNR and transmitter/receiver array sizes.  The number of subcarriers is fixed to $N_{\rm s}=100$ and the number of training symbols is fixed at $\T=100$.  Then, the SNR is varied and the RMSE is calculated and compared to the CRB bound.  Two receiver/transmitter array element numbers are used: $\{N_{\rm rx}=65,N_{\rm tx}=101\}$ and $\{N_{\rm rx}=1,N_{\rm tx}=21\}$.

Fig.~\ref{fig:SNR} compares the simulated RMSE to the CRB bound for each of the channel parameters and path gains for both paths in Fig.~\ref{fig:SNRpaths}.  The legend uses $\{N_{\rm rx},N_{\rm tx}\}$ to convey the transmitter/receiver array sizes.  Each of these plots show that the MSVD channel parameter estimation technique closely matches the CRB bound.  Additionally, it is seen that the larger set of transmitter/receiver antenna elements provides orders of magnitude of improvement in parameter estimation.  Comparison between (a) and (b) in Fig.~\ref{fig:SNR} also shows more accurate AOD estimates than AOA estimates, since the transmitter has more antenna elements than the receiver.

\subsection{Training Symbol Length}

\begin{figure}[t]
	\includegraphics[width=0.49\textwidth]{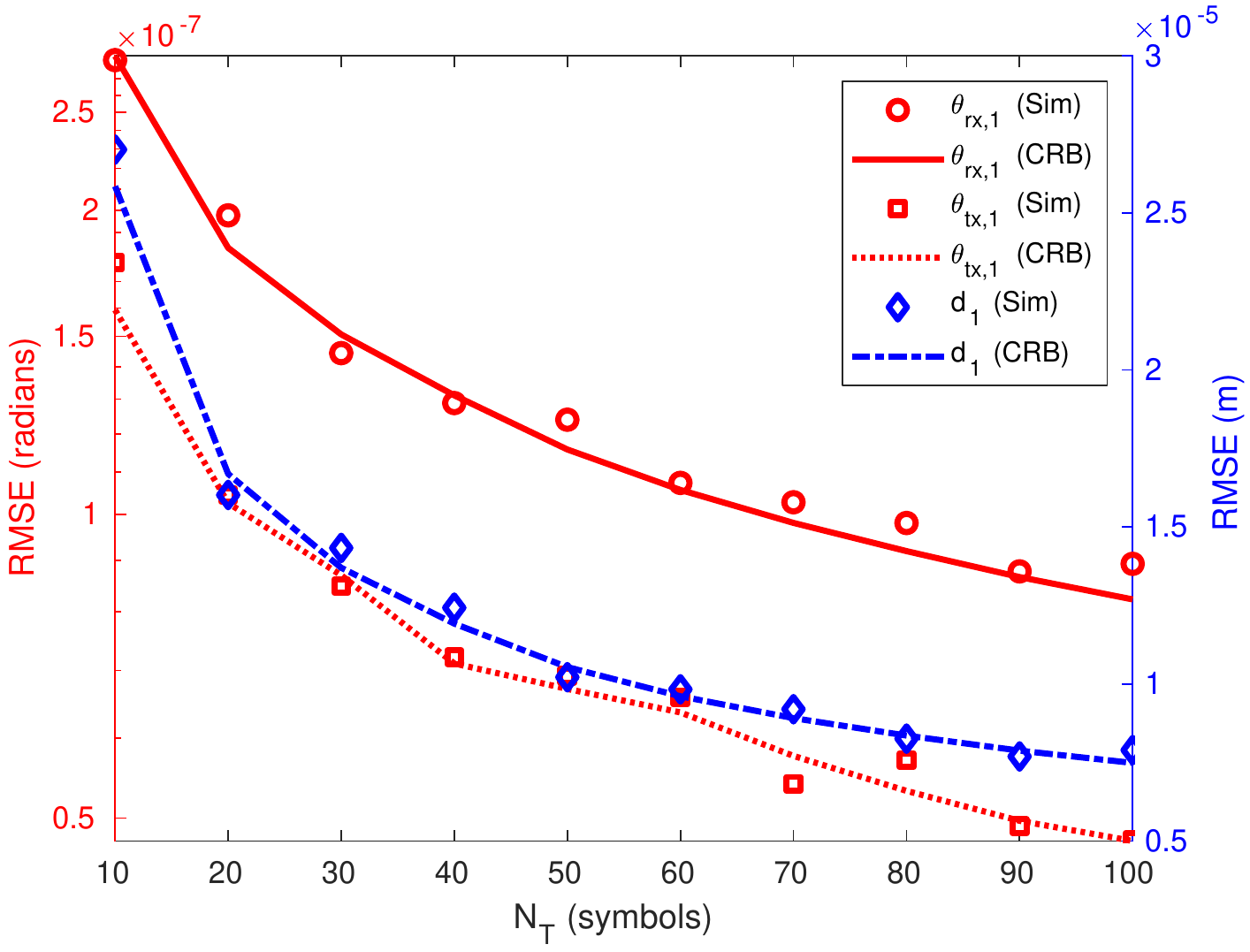}
	\centering
	\caption{Path 1 channel parameters as a function of training symbol length ($\T$).}
	\label{fig:Tsim}
\end{figure}

The number of training symbols ($\T$) is varied with the SNR fixed at $50$ dB and $N_{\rm s}=50$ in Fig.~\ref{fig:Tsim}, which shows the channel parameters from path $1$ only for simplicity.  The RMSE for path $2$ has similar behavior.  It is seen that longer training symbols improve parameter estimation with diminishing improvement as the symbol length increases.  It is also observed that the CRB bound is not smooth as a function of the number of training symbols.  This is because each additional training symbol adds a random column to the measurement tensor in the row subspace.  While the improvement with symbol length is monotonically decreasing, the amount each symbol improves the estimate depends on how independent the extra columns are from the previous columns.  It is also seen that high resolution estimates are obtained with short training symbols (10 symbols).  The slope of the CRB bound is very steep below 10 symbols and channel estimation performance quickly degrades with decreasing training symbol length.

\subsection{Number of Subcarriers}

\begin{figure}[t]
	\includegraphics[width=0.49\textwidth]{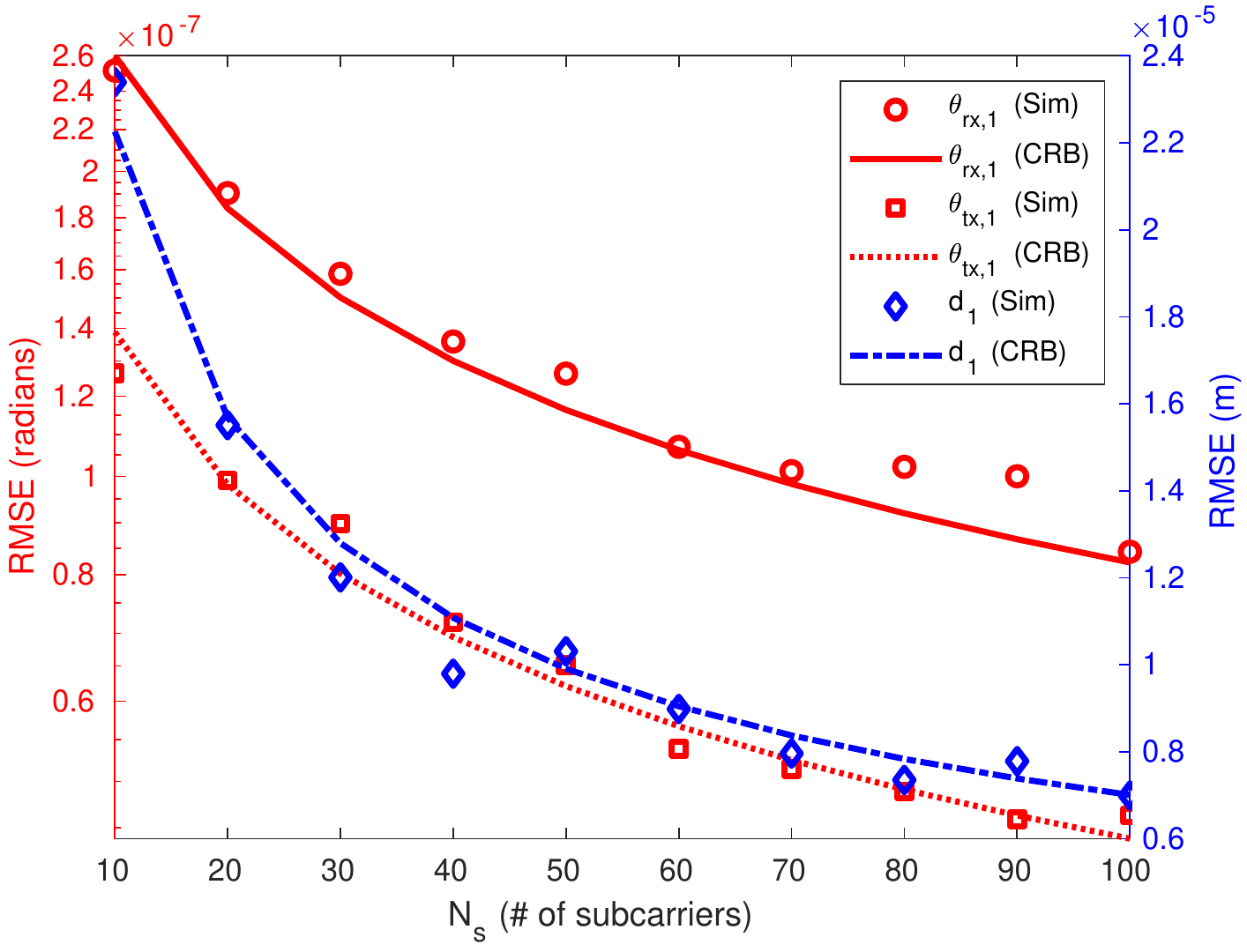}
	\centering
	\caption{Path 1 channel parameters as a function of the number of subcarriers ($N_{\rm s}$).}
	\label{fig:N_s_sim}
\end{figure}

The number of subcarriers ($N_{\rm s}$) is varied with the SNR fixed at $50$ dB and $\T=50$ in Fig.~\ref{fig:N_s_sim}, which also shows the channel parameters from path $1$ only for simplicity.  Channel parameter estimation performance improves with the number of subcarriers as each subcarrier adds a column in the fiber subspace and increases the dimensionality of the measurement tensor.  This also increases the effective SNR.  It is noted that the bandwidth is fixed in this simulation so that the subcarrier spacing deceases with more subcarriers while the OFDM symbol length $\To$ increases.  Allowing the bandwidth and symbol length to both increase leads to even better performance.  

\section{Conclusion}\label{sec:Conclusion}
A channel parameter estimation technique has been proposed based on the MSVD, which is ideally suited for channel parameter estimation since the measurement tensor is naturally represented in Tucker tensor form.  Simulations show that the RMSE obtained using the proposed technique closely matches the CRB. This paper consider scenarios where the transmitter/receiver are limited to a plane.  However, the technique is easily extendable to provide a computationally efficient method for channel parameter estimation in three-dimensional coordinate systems where the elevation angle must also be considered.  Channel parameter estimation has been studied as a function of waveform parameter specifications, where improvements are seen with increased numbers of subcarriers, longer training sequence length, and larger antenna array sizes. Future work will focus on utilizing channel parameter estimation and localization simultaneously for reduced overhead and improved estimation.

\appendices
\section{CRB Derivation for  Parameter Estimation}\label{sec:CRB}
This derivation utilizes the Kronecker product ($\otimes$) and Khatri-Rao product ($\odot$).
The CRB bound is first derived by vectorizing the receiver measurement tensor as shown in \cite{sidiropoulos2017tensor}:
\begin{align}\label{eq:CRBlin}
    \begin{split}
        \vect{\Y} &= \Psib \circledcirc_1 \Wa \circledcirc_2 \Fa \circledcirc_3 \Phib + \N, \\
        &= (\Phib \odot \Fa \odot \Wa)\bm{1} + \vect{\N}, \\
        &= \Xb(\thetab) + \vect{\N},
    \end{split}
\end{align}
where $\Xb(\thetab) = (\Phib \odot \Fa \odot \Wa)\bm{1}$ is a function of the path gains and channel parameters
\begin{equation}
    \thetab = 
    \begin{bmatrix}
        \{\thetarn\}_1^{N_{\rm p}}, & \{\thetatn\}_1^{N_{\rm p}}, & \{d_n\}_1^{N_{\rm p}} & \{h\}_1^{N_{\rm p}}
    \end{bmatrix}.
\end{equation}
The vectorized receiver measurment in \eqref{eq:CRBlin} is in a linear Gaussian form where each of the elements in $\vect{\N}$ are sampled from a $\mathcal{N}(0,\sigma^2)$ and $\sigma^2$ is determined by the SNR.  Therefore, the Fisher information matrix can be calculated as \cite{sidiropoulos2017tensor},\cite{kay1993statistical}:
\begin{equation}
    \bm{I}(\thetab) = \frac{1}{\sigma^2} \bigg[ \frac{\partial \Xb}{\partial \thetab} \bigg]^T \bigg[ \frac{\partial \Xb}{\partial \thetab} \bigg],
\end{equation}
where $\frac{\partial \Xb}{\partial \thetab}$ is the Jacobian of $\Xb$.  

The Jacobian of $\X$ can be divided into four blocks such that
\begin{equation}
    \frac{\partial \Xb}{\partial \thetab} =
    \begin{bmatrix}
        \frac{\partial \Xb}{\partial \bm{\theta}_{rx}} & \frac{\partial \Xb}{\partial \bm{\theta}_{tx}} & \frac{\partial \Xb}{\partial \bm{d}} & \frac{\partial \Xb}{\partial \bm{h}}
    \end{bmatrix}.
\end{equation}
The elements for each block are calculated by rearranging $\Xb$ as shown in \cite{sidiropoulos2017tensor} to the equivalent forms:
\begin{subequations}
\begin{align}
    \Xb &= \big[(\Phib' \odot \Fa) \otimes \bm{I}_{N_{\rm rx}}\big] \vect{\Wa}, \label{eq:Xmana}\\
    &= \bm{K}_{TN_{\rm s},N_{\rm rx}} \big[ (\Wa \odot \Phib') \otimes \bm{I}_T \big] \vect{\Fa}, \label{eq:Xmanb}\\
    &= \bm{K}_{N_{\rm s},N_{\rm rx}T} \big[ (\Fa \odot \Wa') \otimes \bm{I}_{N_{\rm s}} \big] \vect{\Phib'}, \label{eq:Xmanc}
\end{align}
\end{subequations}
where $\bm{K}_{m,n}$ converts the vectorization of the $m \times n$ matrix $S$ to the vectorization of its transpose, or $\bm{K}_{m,n} \vect{\bm{S}} = \vect{\bm{S}^T}$.

The columns of the first block are calculated using \eqref{eq:Xmana} to obtain:
\begin{align}
    \frac{\partial \Xb}{\partial \thetarn} &= \frac{\partial}{\partial \thetarn} \big[(\Phib' \odot \Fa) \otimes \bm{I}_{N_{\rm rx}}\big] \vect{\Wa}, \\
    &= \big[(\Phib' \odot \Fa) \otimes \bm{I}_{N_{\rm rx}}\big] \frac{\partial \vect{\Wa}}{\partial \thetarn}.
\end{align}
The columns of the second block are calculated similarly using \eqref{eq:Xmanb} to obtain:
\begin{equation}
    \frac{\partial \Xb}{\partial \thetatn} = \bm{K}_{TN_{\rm s},N_{\rm rx}} \big[ (\Wa \odot \Phib') \otimes \bm{I}_T \big] \frac{\partial \vect{\Fa}}{\partial \thetatn}
\end{equation}
The columns of the third block are obtained using \eqref{eq:Xmanc}:
\begin{align}
     \frac{\partial \Xb}{\partial d_n} &= \bm{K}_{N_{\rm s},N_{\rm rx}T} \big[ (\Fa \odot \Wa') \otimes \bm{I}_{N_{\rm s}} \big] \frac{\partial \vect{\Phib'}}{\partial \thetatn}.
\end{align}
The columns of the fourth block are calculated using \eqref{eq:Xmana} to obtain:
\begin{align}
    \frac{\partial \Xb}{\partial h_n} &= \bm{K}_{N_{\rm s},N_{\rm rx}T} \big[ (\Fa \odot \Wa') \otimes \bm{I}_{N_{\rm s}} \big] \frac{\partial \vect{\Phib'}}{\partial h_n},
\end{align}
where 
\begin{align}
    \vect{\Phib'} &= \vect{\Phib ~ \text{Diag}(\bm{h}) ~ \bm{I}_{N_{\rm p}}}, \\
    &= (\bm{I}_{N_{\rm p}} \odot \Phib) \bm{h},
\end{align}
where the second step follows from \cite{sidiropoulos2017tensor}.  Then, the fourth block is simplified to:
\begin{equation}
    \frac{\partial \Xb}{\partial h_n} = \bm{K}_{N_{\rm s},N_{\rm rx}T} \big[ (\Fa \odot \Wa') \otimes \bm{I}_{N_{\rm s}} \big] (\bm{I}_{N_{\rm p}} \odot \Phib) \frac{\partial \bm{h}}{\partial h_n}.
\end{equation}

\bibliographystyle{ieeetr} 
\bibliography{overallBib.bib}  

\end{document}